# Tensorial stress-plastic strain fields in α - ω Zr mixture, transformation kinetics, and friction in diamond anvil cell

V. I. Levitas[1,2,3*], Achyut Dhar[1], K. K. Pandey[4]

[1]Department of Aerospace Engineering, Iowa State University, Ames, IA 50011, USA
[2]Department of Mechanical Engineering, Iowa State University, Ames, IA 50011, USA
[3]Ames Laboratory, Division of Materials Science and Engineering, Ames, IA 50011, USA
[4]High Pressure & Synchrotron Radiation Physics Division, Bhabha Atomic Research Centre, Bombay, Mumbai-400085, India

**Abstract:** Various phenomena (phase transformations, chemical reactions, and friction) under high pressures in diamond anvil cell are strongly affected by fields of all components of stress and plastic strain tensors. However, they could not be measured. Even measured pressure distribution contains significant error. Here, we suggest coupled experimental-analytical-computational approaches utilizing synchrotron X-ray diffraction, to solve an inverse problem and find all these fields and friction rules before, during, and after α-ω phase transformation in strongly plastically predeformed Zr. Due to advanced characterization, the minimum pressure for the strain-induced α-ω phase transformation is changed from 1.36 to 2.7 GPa. It is independent of the compression-shear path. The theoretically predicted plastic strain-controlled kinetic equation is verified and quantified. Obtained results open opportunities for developing quantitative high-pressure/stress science, including mechanochemistry, material synthesis, and tribology.

**One Sentence Summary:**
Tensorial stress-plastic strain fields in diamond anvil cell and kinetic rules for α-ω transformation in Zr are determined.

In static high-pressure studies, high pressures are generated by compression, with large elastoplastic deformations, of a thin sample down to 6–20 microns in a diamond anvil cell (DAC) (*1-7*); see **Fig. 1a**. The most advanced characterization of the pressure conditions in a sample is based on determining the radial distribution of pressure averaged over the sample thickness using volume of a crystal cell measured by X-ray diffraction (XRD) and equation of state (EOS) determined under hydrostatic conditions (*4,5,8*). However, EOS for hydrostatic and nonhydrostatic loadings are quite different (*9-12*). More importantly, for the XRD beam along the symmetry axis of the DAC (axial XRD), crystallographic planes that are almost parallel to the beam contribute to the measured XRD patterns only, and axial elastic strain $\bar{E}_{0,zz}$ and consequently stress $\bar{\sigma}_{zz}$ do not contribute to the pressure, leading to large error (bar over the field variables means averaged over the sample thickness). In addition, numerous physical, chemical, geological, and mechanical problems and phenomena are related to knowledge of the fields of all components of the stress, elastic, and plastic strain tensors in materials compressed in DAC (*1-7,9-10,13-28*). For example, contact friction shear stress between diamond and sample/gasket is responsible for generating high pressure and is the key boundary condition for simulation of the processes in DAC (*1,4-6,18,29-37*); however, the friction rules are unknown. It is known that phase transformations (PTs) and chemical reactions strongly depend on the nonhydrostatic stresses and plastic strains (*8,13,17,19-21,23-28,38,39*). New types, namely plastic strain-induced PTs and reactions were formulated and are under intense studies (*8,20,38*). Plastic strain can reduce PT pressure in comparison with quasi-hydrostatic experiment from more than 52.5 to 6.7 GPa for rhombohedral to cubic BN (*24*), from 55 to 5.6 GPa for hexagonal to superhard wurtzitic BN (*17*), and from 70 to 0.7 GPa for graphite to cubic diamond (*19*), which may lead to new technologies. Also, the plastic strain may lead to new nanostructured phases that could not be obtained under hydrostatic conditions, substitute time-dependent kinetics with fast plastic strain-dependent kinetics, and substitute reversible PTs with irreversible PTs that allow one to use retrieved phases in engineering applications (*8,19,20,24,25,38*). Again, since stress and plastic strain tensors are not measurable, quantitative studies of these phenomena are impossible.

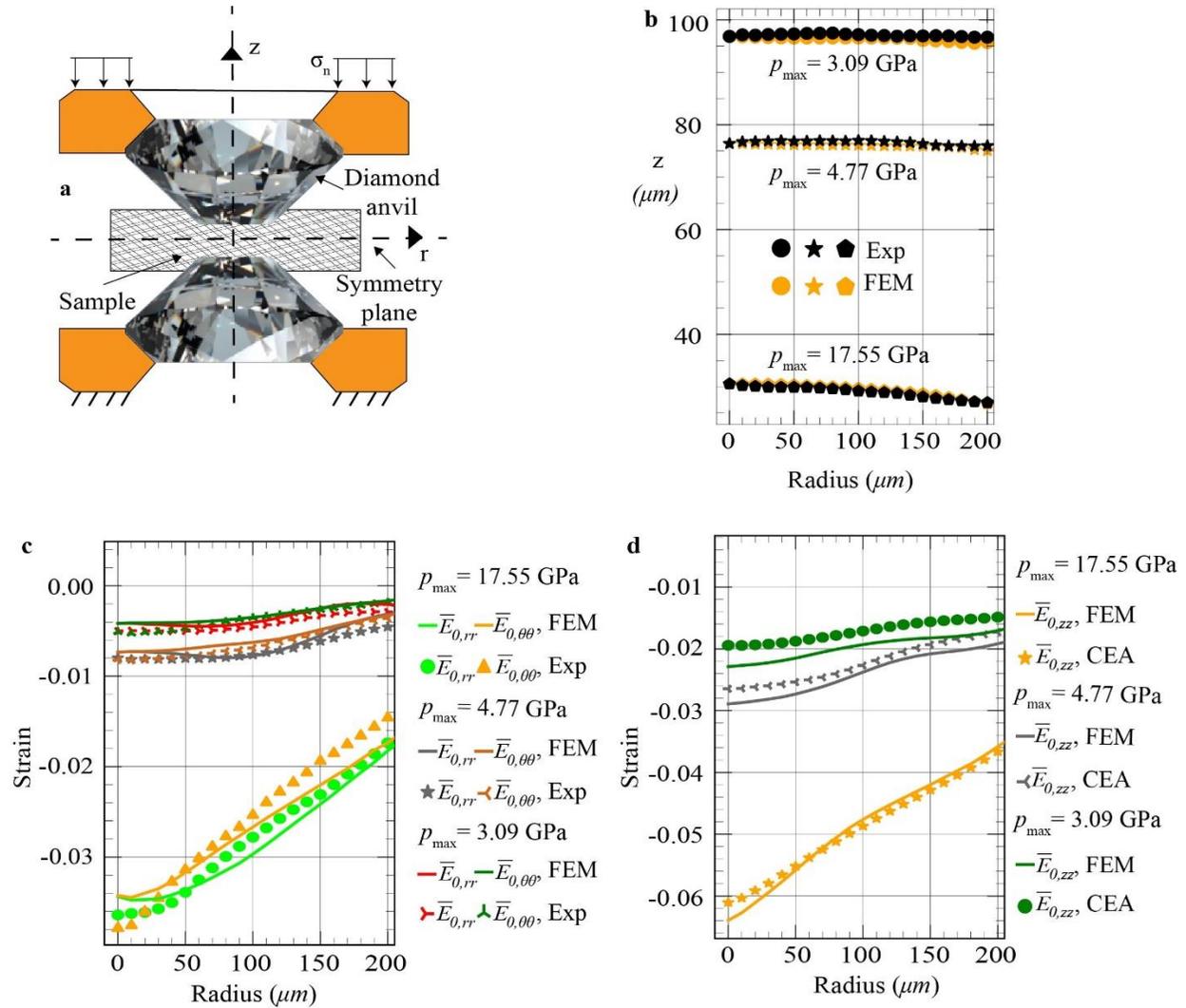

**Fig. 1: Distributions of different strains in DAC.** (a) DAC schematics. (b) The sample thickness profiles from the X-ray absorption and FEM. (c) Comparison of experimental and FEM results for distributions of elastic radial $\bar{E}_{0,rr}(r)$ and the hoop $\bar{E}_{0,\theta\theta}(r)$ strains in a mixture

averaged over the sample thickness. (d) Comparison of FEM and CEA distributions of elastic axial strains $\bar{E}_{0,zz}$.

The only paper that claims measurement of radial distribution of all components of the stress tensor is *(40)*. However, this measurement is performed in a diamond culet rather than in a sample, which gives boundary normal and shear stresses at the sample boundary only instead of full stress and plastic strain tensor fields in the entire sample. Since there was a problem in the precise measurement of the shear stress, finite element method (FEM) simulation was utilized to supplement the experiment. In *(33)*, all tensorial fields are determined in the sample; however, input data are the radial pressure distributions determined using EOS. Measured displacements of material particles at the contact surface with diamond *(35)* represent important boundary conditions, which still were not connected to FEM simulations.

Here, we develop new Coupled Experimental-Analytical (CEA) and CEA-FEM approaches for solving an inverse problem of determining fields of all stress, elastic and plastic strain components (in each phase and in the mixture), and friction shear stress before, during, and after α-ω PT in commercially pure Zr, see flowchart in **Fig. S3**. Importantly, to exclude the effect of strain hardening, change in grain size and dislocation density, and their effect on the thermodynamics and kinetics of PT, we have strongly preliminary deformed Zr until its hardness does not change *(8, 32)*. This is a crucial step to make the problem solvable. Next, based on limited access of the beam for axial geometry, determined fields of XRD patterns, and texture of both phases, we concluded that the most informative and precise approach is to determine, through postprocessing, distributions of the elastic radial $\bar{E}_{0,rr}(r)$ and hoop $\bar{E}_{0,\theta\theta}(r)$ strains in $\alpha$ and $\omega$ phases. Sample thickness profile and pressure-dependence of the yield strength of phases are determined using X-ray absorption *(4, 5, 8)* and broadening of X-ray peaks *(8, 41)*. The CEA approach is developed to determine friction shear stress and all components of stress and elastic strain tensors in each phase and mixture of phases based on XRD measurements. Next, using found friction stress, detailed FEM modeling and simulation are performed that determine all components of stress and plastic strain tensors, which completes the problem in the CEA-FEM approach. Remarkably, the results of analytical and FEM solutions for all stresses are in good correspondence. Distributions of $\bar{E}_{0,rr}(r)$, $\bar{E}_{0,\theta\theta}(r)$ and sample thickness profiles calculated with FEM perfectly correspond to experiments. Obtained pressure distribution differs significantly from that using the EOS-based method. The corrected minimum pressure for the strain-induced α - ω PT is $p_\varepsilon^d$ = 2.70 GPa vs. 1.36 GPa based on the EOS method. Still, it is smaller than under hydrostatic loading by a factor of 2 (and smaller than the phase equilibrium pressure by a factor of 1.3); it is found to be *independent of the plastic strain tensor and its path, in particular, of the compression-shear strain* path. The theoretically predicted plastic strain-controlled kinetic equation was verified and quantified; it is independent of the plastic strain at pressures below $p_\varepsilon^d$ and pressure-plastic strain loading path.

Based on our texture analysis *(42)*, c-axis of α-Zr is predominantly aligned along the loading direction; however, c-axis of ω-Zr is predominantly aligned along the radial direction. In addition, based on the analysis of the XRD peak broadening *(41)*, we obtained for the yield strength of α-Zr $\sigma_y^\alpha = 0.82 + 0.190p$ (GPa) and ω-Zr $\sigma_y^\omega = 1.66 + 0.083p$ (GPa) (**Fig. S2**).

*CEA approach*. Two mechanical equilibrium equations for the axisymmetric model in radial $r$ and axial $z$ directions, the pressure-dependent *von-Mises* yield equation for isotropic perfectly plastic polycrystal $\left(\frac{3}{2}S_{ij}S_{ij}\right)^{0.5} = \sigma_y = (1-c)\sigma_y^\alpha(p) + c\sigma_y^\omega(p)$, and the assumption that radial stress $\sigma_{rr}$ is equal to azimuthal stress $\sigma_{\theta\theta}$, i.e. $\sigma_{rr} = \sigma_{\theta\theta}$, form 4 equations with four unknown stresses, $\sigma_{rr}$, $\sigma_{\theta\theta}$, axial $\sigma_{zz}$, and shear stress $\tau = \tau_{rz}$. Here, $\sigma_y$ is the yield strength in compression of the mixture, $c$ is the volume fraction of ω-Zr, and $S_{ij}$ are components of the deviatoric stress tensor in the mixture. An approximate analytical solution to this statically determined system of equations is found by modifying the Prandtl solution for the plane strain (*42, Section 2*). However, it depends on unknown contact friction shear stress $\tau_c = m\tau_y(p,c)$, where $\tau_c = \frac{\sigma_y(p,c)}{\sqrt{3}}$ is the yield strength in shear of the mixture and $m$ is the factor to be determined. To find the distribution of the friction stress in terms of measured contribution of elastic strain averaged over the sample thickness $\bar{E}_{0,rr}(r) \approx \bar{E}_{0,\theta\theta}(r)$, i.e., in terms of $0.5(\bar{E}_{0,rr}(r) + \bar{E}_{0,\theta\theta}(r))$, the equations derived in (*42, Section 2*) are used.

The modified Hooke's law for hydrostatically pre-stressed properly oriented α- and ω-Zr single crystals with determined pressure-dependent elastic moduli is used to determine stresses in each phase and mixture (*42*). Then both Hooke's law and stress fields from the modified Prandtl solutions are averaged over the sample thickness for each $r$. The *Reuss* hypothesis is used that stresses in a mixture of all α- and ω-Zr single crystals in the representative volume and in polycrystalline aggregate (that participate in the modified Prandtl solution) are the same. Finally, the simplified mechanical equilibrium equation averaged over the sample thickness is utilized, to determine the contact friction shear stress. After the solution of the obtained nonlinear system of algebraic/trigonometric equations for $m(r)$ and $\bar{p}$, all components of the fields of stress and elastic strain tensors in each phase and mixture of phases are obtained analytically, but plastic strains are unknown.

*FEM modeling and simulations (42)*. A large elastoplastic strain model for mixture of α- and ω-Zr using the mixture rule for all properties is advanced. The evolution of the field of the measured volume fraction of ω-Zr $c(r)$ and corresponding isotropic transformation strain are introduced homogeneously along the $z$ coordinate. Obtained analytically evolution of the field $m(r)$ is used as the boundary condition for the contact problem in the culet portion. At the inclined portion of the sample-anvil contact surface, the contact shear stress is determined by the minimum between $\tau_c = m\tau_y(p)$, with the value of $m$ at the culet-inclined surface boundary, and Coulomb friction. The elastic constitutive response of polycrystalline Zr is modeled using 3$^{rd}$ order Murnaghan potential. Associated flow rule in deviatoric stress space is used along with plastic incompressibility. The elastic response of the diamond is modeled using 4$^{th}$ order elastic potential for cubic crystal averaged over azimuthal direction to keep the axial symmetry.

*Results*

Results for three different loadings identified by averaged over the thickness pressure in the mixture $\bar{p}$ at the symmetry axis (i.e., maximum pressure $p_{max}$) are presented in **Fig. 1**. The loadings with $p_{max} = 3.09$ GPa is for the almost pure $\alpha - Zr$ phase, for $p_{max} = 4.77$ GPa is for the mixture of α- and ω-Zr, and for $p_{max} = 17.55$ GPa is for the pure $\omega - Zr$ phase. The corresponding volume fraction profiles $c(r)$ are shown at the bottom of **Fig. 2**. The sample thickness profiles from the X-ray absorption and FEM are in good correspondence (**Fig. 1b**). There is also good correspondence between experiments and FEM distributions of $\bar{E}_{0,rr}(r)$ and $\bar{E}_{0,\theta\theta}(r)$, as well as the closeness of $\bar{E}_{0,rr}(r)$ to $\bar{E}_{0,\theta\theta}(r)$ (**Fig. 1c**). Both experimental verifications of FEM results represent nontrivial validation of the model and simulations; thus, all FEM fields presented below represent reality, even if they cannot be directly measured. **Fig. 1d** shows good correspondence between FEM and CEA distributions of the elastic axial strains $\bar{E}_{0,zz}(r)$, which is a part of the validation of the analytical model.

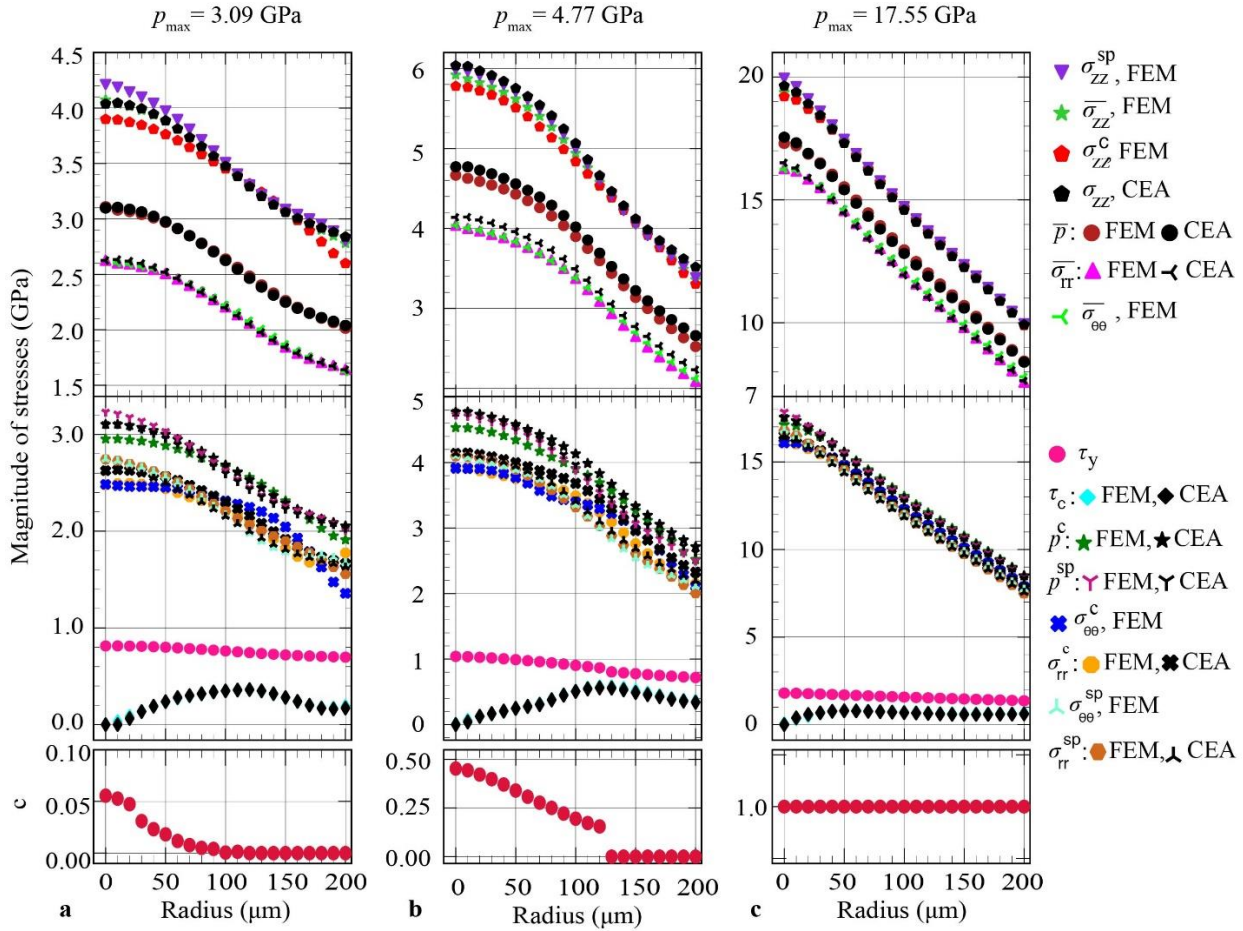

**Fig. 2: Comparison of different radial stress distributions obtained with FEM and analytically.** Results are presented for almost pure $\alpha - Zr$ at $p_{max} = 3.09$ GPa (left column), mixture of α- and ω-Zr for $p_{max} = 4.77$ GPa (middle column), and pure $\omega - Zr$ at $p_{max} = 17.55$ GPa (right column). Distributions of the volume fraction $c(r)$ for the corresponding loadings are shown at the bottom of the figure. Subscripts $'c'$ and $'sp'$ mean contact surface and symmetry plane, respectively.

The comparison of different radial stress distributions obtained with FEM and CEA is given in **Fig. 2**. The practical coincidence of the contact friction stress $\tau_c$ from CEA and FEM is not surprising because the field *m(r)* from the analytical solution is used in FEM as the boundary condition. All stresses smoothly increase from the edge of a culet to the sample center. For all three loadings, there is a very good, and for some stresses, excellent correspondence between the analytical and the FEM results. In addition, FEM distributions of $\sigma_{zz}^c$ at the contact surface and $\sigma_{zz}^{sp}$ at the symmetry plane do not differ essentially, which supports the assumption in the analytical model that $\sigma_{zz}$ is independent of *z*. It can also be seen that $\sigma_{rr} \approx \sigma_{\theta\theta}$ from FEM at the symmetry plane, contact surface, and averaged over the thickness, which justifies the assumption $\sigma_{rr} = \sigma_{\theta\theta}$ made in the analytical model. Comparison of 2D stress contours obtained with CEA and FEM approaches is presented in **Fig. S11**. Maximum values of the same stresses are practically the same, character of changes of all stresses is also the same, i.e., correspondence is good. Thus, despite the simplicity and numerous assumptions, the analytical model describes well stress fields from FEM, and can be used for analysis and interpretation of experiments. Friction shear stress is essentially lower than the yield strength for all pressures. That means that the known method (*18, 29, 30*) to determine the yield strength in shear based on the equilibrium Eq. (S.59) in (*42, Section 2*) and assumption *m=1* does not work. From the edge toward the center, friction stress grows, reaches the maximum and then reduces to zero at the center of a sample due to symmetry conditions.

Friction laws can be formalized with the following equations (**Fig. S10** in (*42*)):

$$\left(\frac{\tau_c}{\tau_y^c}\right)_\omega = 0.186 + 0.018\, p^c \qquad \text{for } 11.1 \leq p^c(\text{GPa}) \leq 15.0; \qquad 60 \leq r\ (\mu m) \leq 140;$$

$$\left(\frac{\tau_c}{\tau_y^c}\right)_{\alpha+\omega} = -0.179 + 0.241\, p^c \qquad \text{for } 2.7 \leq p^c(\text{GPa}) \leq 3.7; \qquad 130 \leq r\ (\mu m) \leq 200; \quad (1)$$

$$\left(\frac{\tau_c}{\tau_y^c}\right)_\alpha = -1.282 + 0.722\, p^c \qquad \text{for } 2.0 \leq p^c(\text{GPa}) \leq 2.45; \qquad 130 \leq r\ (\mu m) \leq 190.$$

The radial distributions of averaged through thickness pressure in α-Zr $\bar{p}^\alpha$ are shown in **Fig. 3a**. They are obtained using developed approach (top) and using hydrostatic (bottom), for the same sample thicknesses. It is evident that the suggested CEA approach to post-process X-ray measurements leads to pressures higher by a factor of 1.7-2.0 or by 1.4-2.0 GPa (i.e., at the level of the yield strength $\sigma_y$ at the corresponding pressure) than those obtained by traditional utilization of EOS. This is a quite significant correction that should be applied to all previous publications in pressure measurements based on EOS (*4, 5*).

The obtained corrections also lead to a reinterpretation of the kinetics of α - ω PT in comparison with that in (*8*). The strain-controlled kinetic equation derived in (*20*) and simplified for Zr in (*8*) is

$$\frac{dc}{dq} = k(1-c)\frac{p(q)-p_\varepsilon^d}{p_h^d - p_\varepsilon^d} \quad \to \quad c = 1 - \exp\left(\frac{-k}{(p_h^d - p_\varepsilon^d)} \int (p(q) - p_\varepsilon^d) dq\right) \qquad (2)$$

where $p$ is pressure either in mixture or in α-Zr ($\bar{p}^\alpha$), $q$ is the accumulated plastic strain, $p_h^d$ is the pressure for initiation of pressure-induced PT under hydrostatic loading, $p_\varepsilon^d$ is the minimum pressure for initiation of the plastic strain-induced PT, $p(q)$ is the loading path. To quantify Eq. (2), experimental points at the center of the sample are used, where unidirectional compression is realized, $q = ln(h_o/h)$, where $h_o$ and $h$ are the initial and current sample thicknesses at the center. For strongly plastically predeformed Zr we found that $p_h^d = 5.4\ GPa$. **Fig. 3b** shows the experimental loading path $p(q)$ based on different pressures: pressures in the parent α-Zr obtained using an analytical model and EOS, as well as averaged over the mixture pressure obtained with an analytical model. The loading path with the analytical model is shifted up with respect to the EOS-based model by 1.3-2.0 GPa. To detect the initiation of PT, three types of markers are superposed on the pressure distribution curves corresponding to $c$=0.05, 0.01, and minimum observable traces of $\omega - Zr$. The minimum pressure for initiation of the plastic strain-induced PT is determined by extrapolating $p$-$c$ results to $c$=0 at the sample center, which gives $p_\varepsilon^d = 2.70$ GPa with CEA approach instead of $p_\varepsilon^d = 1.36$ GPa based on EOS. Thus, the developed method led to an essential increase in the minimum PT pressure. Still, it is 2 times lower than under hydrostatic conditions and lower than the phase equilibrium pressure of 3.4 GPa. Note that with the EOS method, $p_\varepsilon^d$ here for commercially pure Zr is slightly higher than 1.2 GPa for ultra-pure Zr in (*8*). It is important that for both methods of pressure determination, all three types of markers in **Fig. 3a** show very close values for different radii, i.e., the minimum PT pressure is practically independent of $r$. However, the plastic strain tensor and its path are very different for different radii. At the center, unidirectional compression without shear takes place while with increasing radius shear strain grows. Consequently, $p_\varepsilon^d$ is independent of the plastic strain tensor and its path, in particular, of compression/shear plastic strain state and its path. This means that there is no advantage of shear deformation mode in promoting PTs, physical mechanisms are the same for PT under plastic compression and under shear, and PT processes under compression in DAC and torsion in rotational DAC require the same experimental characterization and theoretical treatment. However, rotational DAC allows to independently control pressure and plastic strain and produces PT up to completion close to $p_\varepsilon^d$, which is also important for technologies of plastic strain and defect-induced material synthesis at relatively low pressure.

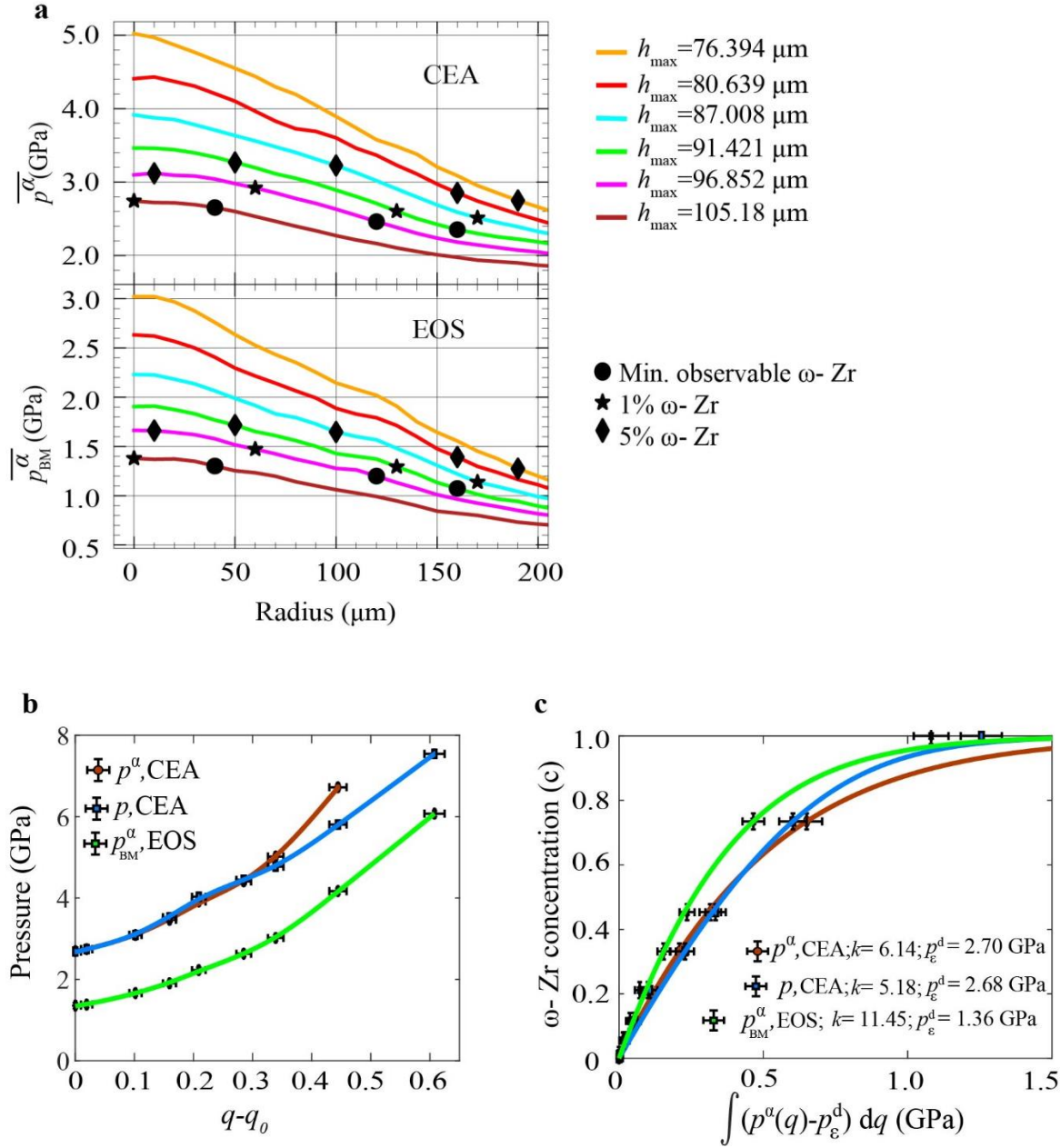

**Fig. 3: Kinetics of α - ω PT.** (a) The radial distribution of $\bar{p}^\alpha$ in α-Zr obtained using developed CEA based on experimentally measured radial $\bar{E}_{0,rr}(r)$ and the hoop $\bar{E}_{0,\theta\theta}(r)$ strains (top) and hydrostatic EOS based on experimentally measured elastic volumetric strain, for the same sample thicknesses. Three types of markers are superposed on these curves corresponding to $c$=0.05, 0.01, and minimum observable traces of $\omega - Zr$. (b) Loading pressure-accumulated plastic strain paths and (c) corresponding kinetics of evolution of $\omega - Zr$ concentration based on pressures in the parent α-Zr obtained using CEA and EOS, as well as averaged over the mixture pressure obtained with the CEA; lines correspond to Eq. (2). Here $q_0$ is the accumulated plastic strain at the beginning of PT.

Experimental points for PT kinetics based on three different pressures are well described by Eq. (2) with different $p_\varepsilon^d$ and kinetic coefficients $k$, see **Fig. 3c**. This validates Eq. (2) for a quite nontrivial loading path. While the difference between kinetic curves with pressures based on the CEA approach and EOS does not look drastic, this is due the choice of independent variables along the horizontal axis, different for different cases. Quantitatively, the CEA method led not only to an increase in $p_\varepsilon^d$ by a factor of 2 but also to a decrease in kinetic coefficient $k$ from 11.45 to 6.14, i.e., correction is very significant. Since this correction did not change the conclusion that $p_\varepsilon^d$ is independent of compression/shear plastic strain state and its path, we assume that other conclusions from (8), that Eq. (2) is independent of the magnitude of plastic strain $q_o$ below $p_\varepsilon^d$ and of $p - q$ loading path, are valid as well.

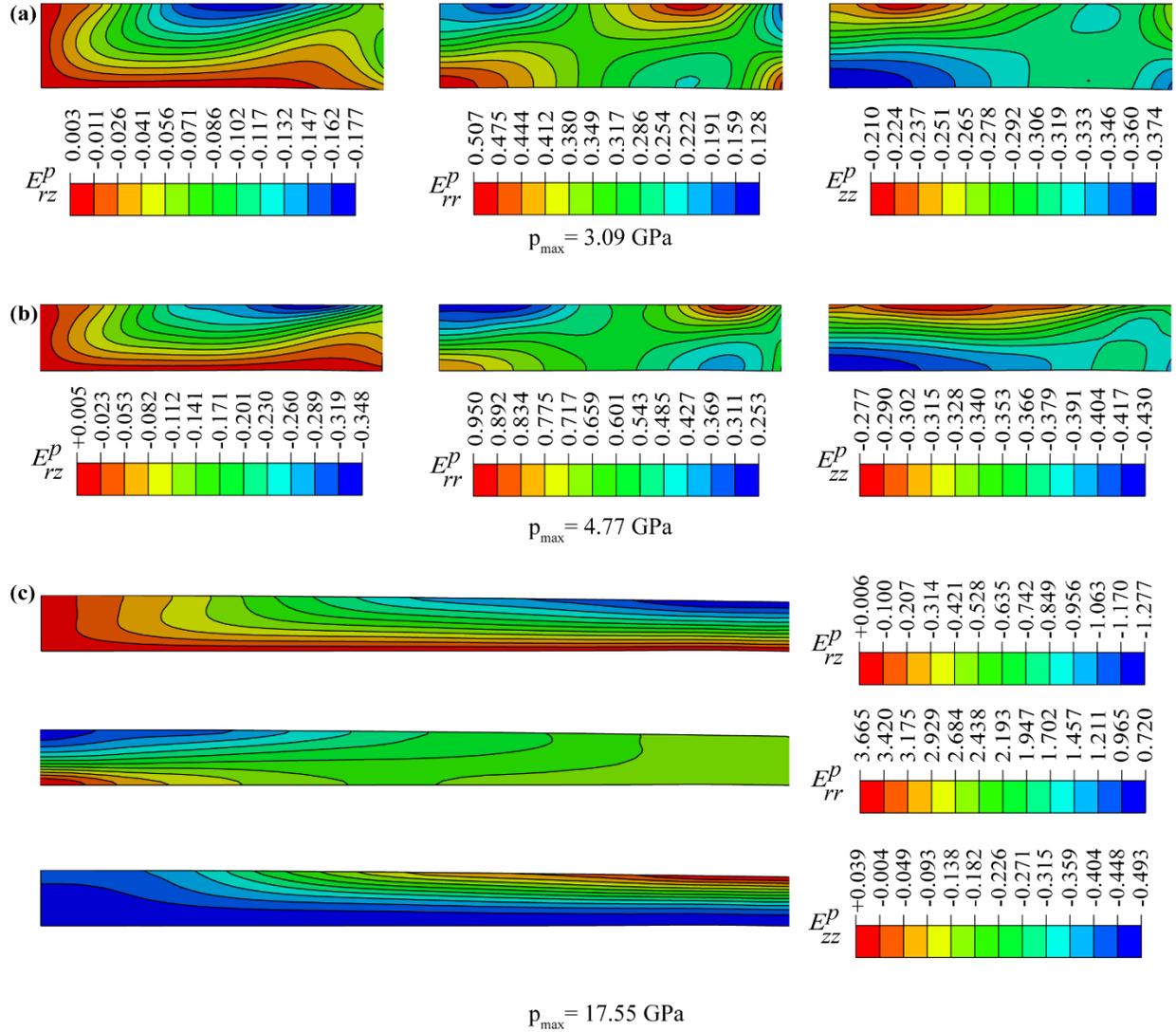

**Fig. 4:** Distributions of components of Lagrangian plastic strains in a sample for three loadings.

Fields of all components of Lagrangian plastic strain tensor found with the CEA-FEM approach are presented in **Fig. 4**. Very heterogeneous and nontrivial distributions are observed, caused by heterogeneous contact friction. These fields will be used in future work, which will include simulation of the strain-induced PT as well, to derive, calibrate, and validate a more precise kinetic equation for strain-induced PT.

To summarize, we developed coupled CEA and CEA-FEM approaches that allow us to solve the inverse problem and reinterpret X-ray diffraction measurements and determine pressure and tensorial stress-plastic strain fields in each phase and mixture, as well as contact friction rules in a DAC before, during, and after α-ω transformation in Zr. Good correspondence of the elastic radial $\bar{E}_{0,rr}(r)$ and hoop $\bar{E}_{0,\theta\theta}(r)$ strains and the sample thickness profile between FEM and experiments and the stress tensor fields between the CEA approach and FEM validates the developed approach. Due to novel characterization, the minimum pressure $p_\varepsilon^d$ for the strain-induced α-ω PT in Zr is changed from 1.36 to 2.7 GPa and the kinetic coefficient $k$ is reduced from 11.45 to 6.14, i.e., correction is very significant. Still, $p_\varepsilon^d$ is 2 times lower than under hydrostatic conditions and lower than the phase equilibrium pressure of 3.4 GPa. The found independence of $p_\varepsilon^d$ of the plastic strain tensor and its path (in particular, compression-shear path) means basic equivalence of PT processes under compression in DAC and torsion in rotational DAC. However, rotational DAC allows to independently control pressure and plastic strain and produces PT up to completion at pressures close to $p_\varepsilon^d$, which is important for technologies of plastic strain and defect-induced material synthesis at relatively low pressure. Since our pressure correction did not change the conclusion that $p_\varepsilon^d$ is independent of compression/shear plastic strain state and its path, it is highly probable that other conclusions from (*8*), that kinetic Eq. (1) is independent of the magnitude of plastic strain $q_o$ below $p_\varepsilon^d$ and of *p-q* loading path, are valid as well.

Obtained results open opportunities for developing quantitative high-pressure/stress science. The same methods can be extended for other material systems and processes in rotational DAC. They can significantly improve the accuracy of pressure field determination and characterization of all processes studied under pressure: physical, chemical, biological, geophysical, and others. Finding fields of stress tensor components (which can be done using analytical model) will allow quantifying their effect on the processes under study, instead of referring to the qualitative effect of pressure-transmitting media and "non-hydrostatic" stresses. Finding fields of plastic strain tensor components, which cannot be measured, will allow to quantitatively study plastic strain induced PTs and chemical reactions and initiate quantitative high-pressure mechanochemistry. This may lead to new technologies of plastic strain and defect-induced material synthesis at relatively low pressure, in particular, for diamond (*19*) and cubic BN (*17, 24*), initiation of high-pressure tribology, explanation of deep-focus earthquakes (*26*), the appearance of microdiamond in the low pressure and temperature Earth's crust (*19*), and mechanochemical origin of life in the icy crust of solar system's moons and planets (*27, 28*). Our method generates big data from the single experiments/simulation, which can be utilized for machine learning-based development and calibration of the corresponding constitutive equations. Thus, the main challenge, namely, strong heterogeneity of all fields, can be transformed into an opportunity.

Supplementary materials for

**Stress-plastic strain fields in $\alpha$ - $\omega$ Zr mixture, transformation kinetics, and friction in diamond anvil cell**

V.I. Levitas[1,2,3*], Achyut Dhar[1], K.K. Pandey[4]

[1] Department of Aerospace Engineering, Iowa State University, Ames, IA 50011, USA
[2] Department of Mechanical Engineering, Iowa State University, Ames, IA 50011, USA
[3] Ames Laboratory, Division of Materials Science and Engineering, Ames, IA 50011, USA
[4] High Pressure Synchrotron Radiation Physics Division, Bhabha Atomic Research Centre, Bombay, Mumbai-400085, India

**This PDF file includes:**
Material and Experimental Methods
Pressure -dependence of the yield strengths of $\alpha-$ and $\omega-$Zr
Equation of state under hydrostatic loading
Supplementary text
Figures S1 to S11
Tables S1 to S2
References

**Material and Experimental Methods**

The material studied in the paper is the same as was used by Zhilyaev et al. [1], purchased from Haines and Maassen (Bonn, Germany), i.e., commercially pure (99.8%) $\alpha-Zr$ (Fe: 330 ppm; Mn: 27 ppm; Hf: 452 ppm; S: <550 ppm; Nd: < 500 ppm). The sample slab of initial thickness of 5.25 $mm$ was cold rolled down to $\sim$ 165 $\mu m$ to obtain plastically pre-deformed sample with saturated hardness. Vickers microhardness test method was used to characterize hardness of sample at several steps during cold rolling. A 3 $mm$ diameter disk was punch cut from thus obtained thin rolled sheet for unconstrained compression experiments in diamond anvil cell (DAC). For hydrostatic compression experiments, small specks of $\sim$ 20 $\mu m$ size were chipped off from the plastically pre-deformed sample using diamond file.

The hydrostatic high-pressure X-ray diffraction measurements were carried out using same DAC to estimate equation of state, bulk modulus, and its pressure derivative at ambient pressure for this sample. For these experiments, small Zr specks of $\sim$ 20 $\mu m$ size, as already mentioned, were loaded in sample chamber along with silicone oil and copper chips as pressure transmitting medium and pressure marker respectively. The sample chamber was prepared by drilling a hole of $\sim$ 250 $\mu m$ diameter in steel gaskets pre-indented using diamond anvils from initial thickness of $\sim$ 250 $\mu m$ to $\sim$ 50 $\mu m$. Hydrostatic high-pressure experiments were carried out in small pressure steps of $\sim$ 0.2 GPa up to a maximum pressure of 16 GPa.

Unconstrained plastic compression experiments were carried out prescribing different compression loads to plastically pre-deformed Zr sample loaded in DAC without any constraining gasket. The sample was subjected to axial loads of 50 N, 100 N, 150 N, 170 N, 190 N, 210 N, 230 N, 250 N, 270 N, 290 N, 310 N, 330 N, 350 N, 400 N, 450 N, 500 N, 550 N, 600 N, 650 N, 700 N, 750 N, 800 N, 850 N, 900 N, 950 N and 1000 N.

In-situ XRD experiments were performed at 16-BM-D beamline at HPCAT sector at Advanced Photon Source employing focused monochromatic X-rays of wavelength 0.3096(3)Å and size $\sim 6\mu m \times 5\mu m$ (full width at half maximum (FWHM)). At each load-condition, the sample was radially scanned over the entire culet diameter (500 $\mu m$) in steps of 10 $\mu m$ and 2D diffraction images were recorded at Perkin Elmer flat panel detector. At each load step, X-ray absorption scan was also recorded in same 10 $\mu m$ steps to obtain thickness profile of sample under given load condition.

2D diffraction images were converted to a 1D diffraction pattern using FIT2D software [2, 3] and subsequently analyzed through Rietveld refinement [4, 5] using GSAS II [6] and MAUD [7] software for obtaining lattice parameters, phase fractions and texture parameters of both $\alpha$ and $\omega$ phases of Zr.

In axial geometry (i.e., when the incident X-ray beam is directed along z-axis) (Fig. S1 and Fig. 1a in the main text), the diffraction condition is satisfied mostly for those crystallographic planes which are nearly parallel (plane normal perpendicular) to the load axis. Hence the observed shifts in diffraction peaks can be practically used to estimate strains in radial and azimuthal directions viz. $\bar{E}_{0,11} = \bar{E}_{0,rr}$ and $\bar{E}_{0,22} = \bar{E}_{0,\theta\theta}$ averaged over the sample thickness. Ideally, the angle between the load axis and diffraction vector $\psi$ should be equal to 90° to estimate these strain components. However, since this is not possible in axial geometry, we can use the diffraction peak with smallest diffraction angle, $\theta$. In our experiments for $\alpha$-Zr, (100) diffraction peak appears at $\theta = 3.18°$ for used X-rays ($\lambda = 3.1088$Å) at ambient pressure. This corresponds to $\psi = 86.82°$ and can be used for estimation of strain components $\bar{E}_{0,rr}$ and $\bar{E}_{0,\theta\theta}$. Note that (100) peak corresponds to $'a'$ lattice parameter because c-axis of $\alpha$-Zr is predominantly aligned along the loading direction as per our texture analysis.

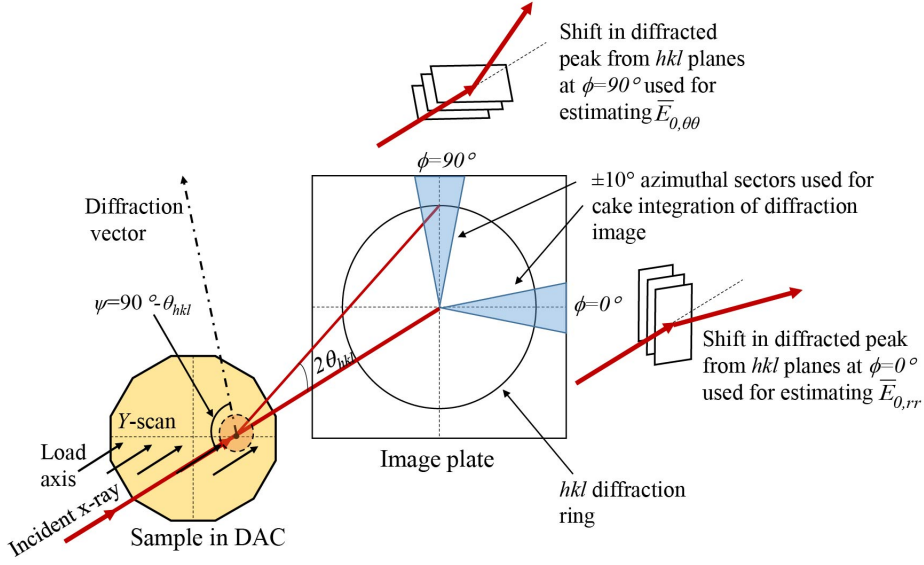

Figure S1: **Schematic illustration of estimation of elastic strains in radial $\bar{E}_{0,rr}$ and azimuthal $\bar{E}_{0,\theta\theta}$ directions at each scanning position.**

For $\omega$-Zr, (001) diffraction peak appearing at $\theta = 2.85°$ ($\psi = 87.15°$) can be used for estimation of strain components $\bar{E}_{0,rr}$ and $\bar{E}_{0,\theta\theta}$. The (001) peak of $\omega$-Zr corresponds to $'c'$ lattice parameter and as per texture analysis, c-axis of $\omega$-Zr is predominantly perpendicular to the loading direction of DAC.

Thus, strain components $\bar{E}_{0,rr}$ and $\bar{E}_{0,\theta\theta}$ for $\alpha$ and $\omega$ phases of Zr have been obtained for each loading condition at each scanning position using the following equations:

For $\alpha$-Zr:
$\bar{E}_{0,rr} = 0.5((a/a_0)^2 - 1)$ using $\phi = 0°$ sector of (100) diffraction ring;
$\bar{E}_{0,\theta\theta} = 0.5((a/a_0)^2 - 1)$ using $\phi = 90°$ sector of (100) diffraction ring;

For $\omega$ -Zr :
$\bar{E}_{0,rr} = 0.5((c/c_0)^2 - 1)$ using $\phi = 0°$ sector of (100) diffraction ring;
$\bar{E}_{0,\theta\theta} = 0.5((a/a_0)^2 - 1)$ using $\phi = 90°$ sector of (100) diffraction ring;

Finally, diffraction data at the symmetry axis for all load conditions were used for quantitative analysis of the kinetics of plastic-strain induced $\alpha - \omega$ phase transition in Zr, like in [8]. For this purpose, the pressure in $\alpha$-Zr and volume fraction of $\omega$-Zr were estimated as a function of accumulated plastic strain $q$. At the symmetry axis, material experiences an unidirectional compression, for which $q = ln(h_0/h)$, where $h_o$ is the initial thickness of the sample in DAC and $h$ is the current thickness.

It is important to note that the directions 1, 2, and 3 coincide with the radial, azimuthal, and $z$ directions, respectively, see Fig. S5 below. This direct correspondence between designations applies to components of all tensors.

**Pressure-dependence of the yield strengths of α- and ω-Zr**

The yield strengths of $\alpha$ and $\omega$ phases of Zr were estimated using the peak broadening method [9] near the center of a sample (see Fig. S2):

$$\sigma_y^\alpha = 0.82 + 0.19p \text{ (GPa)} \quad \text{and} \quad \sigma_y^\omega = 1.66 + 0.083p \text{ (GPa)}$$

It is worth mentioning here that Zhao et al. [9] reported the yield strengths of $\alpha$ and $\omega$ phases as 0.18 GPa and 1.18 GPa at ambient pressure, respectively. The reason that our values of the yield strengths are significantly higher is that our Zr sample was subjected to large preliminary plastic deformation until saturation of the strain hardening, while Zhao et al. performed experiments on an annealed Zr sample.

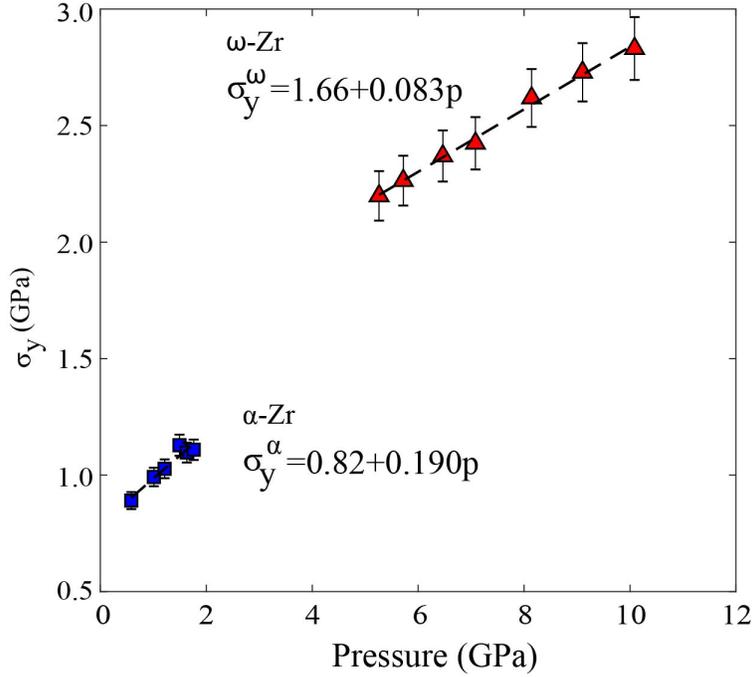

Figure S2: **Pressure dependence of yield strength in compression of α- and ω-Zr.**

**Equation of state under hydrostatic loading**

Experiments under hydrostatic loading were performed for comparison and to determine the equation of state for both phases. The $\alpha \to \omega$ PT started at pressure 5.4 GPa and finished at 6.6 GPa. The 3rd order Birch-Murnaghan equation of state (EOS) fitting of pressure-volume data for $\alpha$ and $\omega$ phases of Zr provides: initial (ambient) volumes $V_0 = 23.269 \text{Å}^3$ and $34.306 \text{Å}^3$ (per formula unit); bulk moduli $K_0 = 93.55$ GPa and 102.4 GPa @ $p = 0$, and pressure derivative of bulk moduli $K'=3.0$ and 2.93 @ $p = 0$, respectively. $\omega$-Zr is retained at ambient pressure on complete pressure release. Components of the deformation gradient $\boldsymbol{F}_*(p)$ determined in the hydrostatic experiments for $\alpha$ and $\omega$ phases are presented in Eqs. (S.4)- (S.7).

# Supplementary text

## 1 Coupled experimental-analytical-computational approaches for finding stress and plastic strain tensor fields and friction rules in a sample compressed in DAC

The flowchart of the interaction between different methods is presented in Fig. S3.

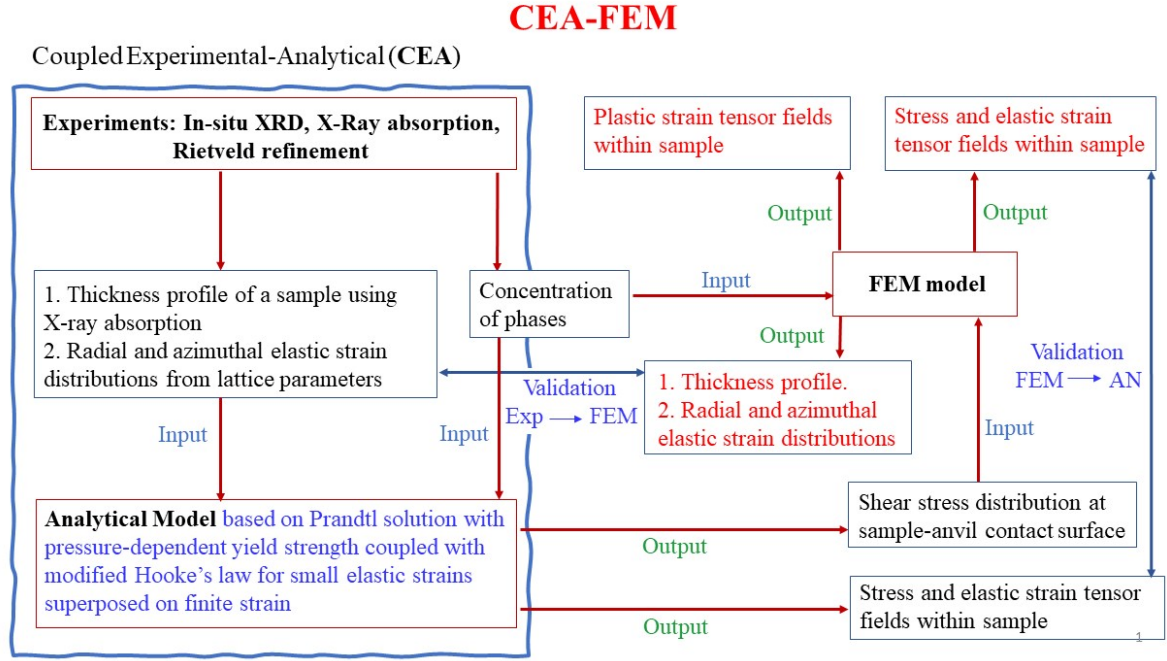

Figure S3: **The flowchart of the interaction between experimental, analytical, and FE methods.**

Experimental methods based on in-situ X-ray diffraction and absorption allow us to find the radial $\bar{E}_{0,rr}$ and the azimuthal $\bar{E}_{0,\theta\theta}$ strain distributions in $\alpha$-Zr and $\omega$-Zr phases, and concentration of $\omega$ phase $c(r)$ (all averaged over the sample thickness), as well as sample thickness profile $h(r)$. These distributions are the input data for our analytical model, which allows us to determine distribution of the contact friction stress at the sample-anvil surface and 2D fields of all components of the stress and elastic strain tensors in each phase and mixture. The friction stress distribution is utilized as the boundary condition in our FEM problem formulation; evolution of concentration of $\omega$ phase $c(r)$ is introduced homogeneously along the z-coordinate in our FEM problem formulation. FEM solution delivers all components of stress, elastic, and plastic strain tensors and the sample thickness profile. FEM-based radial $\bar{E}_{0,rr}$ and azimuthal $\bar{E}_{0,\theta\theta}$ strain distributions and the sample thickness profile are compared with experiments to validate FEM modeling, and, consequently, the entire procedure and all fields. All components of the stress and elastic strain tensors in the mixture from the analytical solution are compared with FEM solution to validate the analytical model.

## 2 Coupled Experimental-Analytical (CEA) approach

To obtain the analytical solution for stress and elastic strain fields in a sample compressed in DAC, we will make number of strong and counterintuitive assumptions. Surprisingly, the final analytical solution is in good agreement with much more precise FEM solution (Fig. 1 and Fig. 2 in the main text), which justifies the admissibility of our assumptions.

### 2.1 Modified Hooke's law under pressure [10]

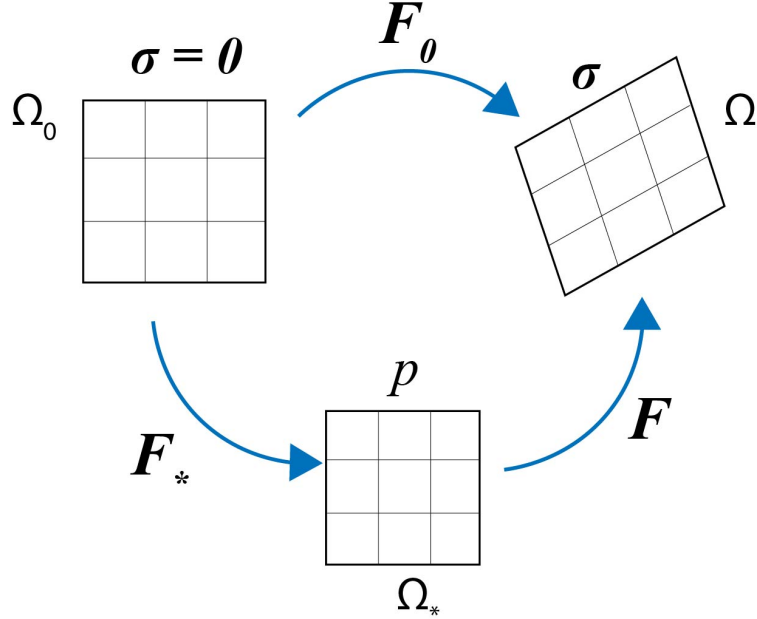

Figure S4: **The stress-free configuration $\Omega_0$, the deformed (current) configuration $\Omega$ under the Cauchy stress $\sigma$, and an arbitrary intermediate configuration $\Omega_*$ with Cauchy stress $\sigma_*$. Multiplicative decomposition of the deformation gradient $\boldsymbol{F}_0 = \boldsymbol{F} \cdot \boldsymbol{F}_*$ is valid.**

Let the total deformation gradient from the undeformed stress-free reference configuration to the current deformed configuration $\boldsymbol{F}_0 = \boldsymbol{F} \cdot \boldsymbol{F}_*$ is decomposed multiplicatively into deformation gradient $\boldsymbol{F}_*$ corresponding to hydrostatically loaded with pressure $p$ intermediate configuration and the deformation gradient $\boldsymbol{F}$ that describes small strains from the intermediate to current configuration (Fig. S4). Then the modified Hooke's law is:

$$\boldsymbol{\sigma} = -p(\boldsymbol{F}_*)\boldsymbol{I} + \boldsymbol{B}(p) : \tilde{\boldsymbol{\varepsilon}}; \qquad \tilde{\boldsymbol{\varepsilon}} = (\boldsymbol{F} - \boldsymbol{I})_s; \tag{S.1}$$

$$\sigma_{ij} = -p(F_{*mn})\delta_{ij} + B_{ijkl}(p)\tilde{\varepsilon}_{lk}; \qquad \tilde{\varepsilon}_{lk} = (F_{lk} - \delta_{lk})_s, \tag{S.2}$$

where $\boldsymbol{\sigma}$ is the Cauchy (true) stress, pressure $p = -(\sigma_{11} + \sigma_{22} + \sigma_{33})/3$, $\boldsymbol{I}$ is the identity tensor, $\boldsymbol{B}$ is the pressure-dependent elastic moduli tensor that connects Jaumann derivative of the Cauchy stress and the strain-rate (often called the Wallace moduli). Deformation gradient $\boldsymbol{F}_*(p)$ is measured under hydrostatic DAC experiments and for hexagonal crystals

considered here it is:

$$\boldsymbol{F}_*(p) = \begin{bmatrix} F_{*11} & 0 & 0 \\ 0 & F_{*22} & 0 \\ 0 & 0 & F_{*33} \end{bmatrix}, \tag{S.3}$$

where $F_{*11}(p) = F_{*22}(p) = \frac{a_*}{a_0}$ and $F_{*33}(p) = \frac{c_*}{c_0}$, $a_0, c_0$ and $a_*, c_*$ are the lattice parameters in the reference and pressurized configurations, respectively. It is important to note that here $\boldsymbol{F}_*$ is written in the local crystal coordinate system, i.e. $F_{*11}$ and $F_{*22}$ are in the basal slip plane, and $F_{*33}$ is normal to the basal slip plane.

Results of our hydrostatic experiments can be approximated as

$$F_{*11}^\alpha(p) = 1.00 - 0.00372p + 0.00006p^2; \tag{S.4}$$

$$F_{*33}^\alpha(p) = 1.0 - 0.00333p + 0.00006p^2; \tag{S.5}$$

$$F_{*11}^\omega(p) = 1.00 - 0.00330p + 0.00003p^2, \tag{S.6}$$

and

$$F_{*33}^\omega(p) = 1.0 - 0.00282p + 0.00002p^2. \tag{S.7}$$

Approximation for $\alpha - Zr$ is valid up to 10 GPa, and for $\omega - Zr$ up to 20 GPa.

Since $\boldsymbol{F}_*$ is diagonal and $\boldsymbol{F}$ describes small strains, then $\boldsymbol{F}_0$ has the form

$$\boldsymbol{F}_0 = \begin{bmatrix} F_{0,11} & \gamma_1 & 0 \\ \gamma_2 & F_{0,22} & 0 \\ 0 & 0 & F_{0,33} \end{bmatrix}, \tag{S.8}$$

where generally $\gamma_1 \neq \gamma_2$ because of rotations superposed on the shear strains, and we took into account zero shears for axisymmetric problem.

We need to express $\tilde{\boldsymbol{\varepsilon}}$, which participates in the Hooke's law (S.1), with $\boldsymbol{F}$ and $\boldsymbol{F}_*$, which are measured. Since $\tilde{\boldsymbol{\varepsilon}}$ is small, therefore,

$$\tilde{\boldsymbol{\varepsilon}} \approx \boldsymbol{E} = \boldsymbol{F}_*^{-1T}(p) \cdot (\boldsymbol{E}_0 - \boldsymbol{E}_*(p)) \cdot \boldsymbol{F}_*^{-1}(p); \tag{S.9}$$

$$\tilde{\varepsilon}_{lk} \approx E_{lk} = F_{*ik}^{-1}(p)(E_{0,ij} - E_{*ij}(p))F_{*jl}^{-1}(p), \tag{S.10}$$

where $\boldsymbol{E} = 0.5\left(\boldsymbol{F}^T \cdot \boldsymbol{F} - \boldsymbol{I}\right)$ is the Lagrangian strain corresponding to $\boldsymbol{F}$, $\boldsymbol{E}_* = 0.5\left(\boldsymbol{F}_*^T \cdot \boldsymbol{F}_* - \boldsymbol{I}\right)$ and $\boldsymbol{E}_0 = 0.5\left(\boldsymbol{F}_0^T \cdot \boldsymbol{F}_0 - \boldsymbol{I}\right)$ are the Lagrangian strains corresponding to $\boldsymbol{F}_*$ and $\boldsymbol{F}_0$. For

$\boldsymbol{F}_0$ in Eq. (S.8), we obtain

$$\boldsymbol{E}_0 = 0.5 \begin{bmatrix} F_{0,11}^2 - 1 + \gamma_2^2 & F_{0,11}\gamma_1 + F_{0,22}\gamma_2 & 0 \\ F_{0,11}\gamma_1 + F_{0,22}\gamma_2 & F_{0,22}^2 - 1 + \gamma_1^2 & 0 \\ 0 & 0 & F_{0,33}^2 - 1 \end{bmatrix} \simeq$$

$$0.5 \begin{bmatrix} F_{0,11}^2 - 1 & \gamma & 0 \\ \gamma & F_{0,22}^2 - 1 & 0 \\ 0 & 0 & F_{0,33}^2 - 1 \end{bmatrix}; \qquad \gamma := F_{0,11}\gamma_1 + F_{0,22}\gamma_2, \qquad (S.11)$$

where we neglected small $\gamma_i^2$ in comparison with the finite $F_{0,ii}^2 - 1$. Similarly,

$$\boldsymbol{E}_* = 0.5 \begin{bmatrix} F_{*11}^2 - 1 & 0 & 0 \\ 0 & F_{*22}^2 - 1 & 0 \\ 0 & 0 & F_{*33}^2 - 1 \end{bmatrix}. \qquad (S.12)$$

Then, based on Eq. (S.9),

$$\tilde{\boldsymbol{\varepsilon}} = 0.5 \begin{bmatrix} \frac{E_{0,11} - E_{*11}}{F_{*11}^2} & \frac{\gamma}{F_{*11}F_{*22}} & 0 \\ \frac{\gamma}{F_{*11}F_{*22}} & \frac{E_{0,22} - E_{*22}}{F_{*22}^2} & 0 \\ 0 & 0 & \frac{E_{0,33} - E_{*33}}{F_{*33}^2} \end{bmatrix} = 0.5 \begin{bmatrix} \frac{F_{0,11}^2}{F_{*11}^2} - 1 & \frac{\gamma}{F_{*11}F_{*22}} & 0 \\ \frac{\gamma}{F_{*11}F_{*22}} & \frac{F_{0,22}^2}{F_{*22}^2} - 1 & 0 \\ 0 & 0 & \frac{F_{0,33}^2}{F_{*33}^2} - 1 \end{bmatrix}. \qquad (S.13)$$

With the obtained structure of $\tilde{\boldsymbol{\varepsilon}}$, the Hooke's law for a hydrostatically pre-stressed hexagonal crystals can be split for normal and shear $\tau$ stresses:

$$\begin{bmatrix} \sigma_{11} + p \\ \sigma_{22} + p \\ \sigma_{33} + p \end{bmatrix} = \begin{bmatrix} \hat{\sigma}_{11} \\ \hat{\sigma}_{22} \\ \hat{\sigma}_{33} \end{bmatrix} = \begin{bmatrix} B_{11} & B_{12} & B_{13} \\ B_{12} & B_{22} & B_{23} \\ B_{13} & B_{23} & B_{33} \end{bmatrix} \begin{bmatrix} \tilde{\varepsilon}_{11} \\ \tilde{\varepsilon}_{22} \\ \tilde{\varepsilon}_{33} \end{bmatrix}; \qquad \tau = \tau_{13} = 0.5 B_{44} \frac{\gamma}{F_{*11} F_{*22}}, \qquad (S.14)$$

where the simplifications due to hexagonal symmetry of $B_{ij}$ ($B_{11} = B_{22}$, $B_{13} = B_{23}$, $B_{44} = B_{55}$, and $B_{66} = 0.5(B_{11} - B_{12})$) were not yet applied. Since XRD does not measure elastic shear strains, equation for shear components will not be used in further derivations. However, after we will find field of the shear stress $\tau$ in the sample, we can use the second Eq. (S.14) to find elastic shear stain $\gamma$.

Pressure-dependence of the single crystal elastic moduli $B_{ij}$ was approximated by a quadratic polynomial with parameters given for $\alpha$-Zr in Table S1 and for the effective elastic moduli $B_{ij}^e$ for $\omega$-Zr in Table S2. They were produced by combining published experimental and first-principle results ([11] for $\alpha$-Zr and [12] for $\omega$-Zr) and implementing consistency conditions (see [10]) with our hydrostatic experiments; i.e., elastic moduli $B_{ij}$ reproduce generalized equations of state Eqs. (S.4)-(S.7). Effective elastic moduli are defined in Eq. (S.66) with allowing for an actual orientation of a single crystal with respect to the coordinate system and some additional symmetry requirements.

Note that small strain $\tilde{\boldsymbol{\varepsilon}}$ causes change in pressure. We can iteratively update $p$ (and corresponding $\boldsymbol{F}_*$) in the intermediate configuration, so that it coincides with the pressure in the current configuration. Then strain $\tilde{\boldsymbol{\varepsilon}}$ will produce deviatoric stress, which is limited by the yield strength. Therefore, strain $\tilde{\varepsilon}$ is also limited and is small in comparison with $\boldsymbol{E}_*$, which does not have any constraints.

Table S1: Elastic constants and their pressure derivatives for $\alpha$-Zr that meet the consistency conditions.

| $\alpha$-Zr | $B_{11}$ | $B_{33}$ | $B_{12}$ | $B_{13}$ | $B_{44}$ |
|---|---|---|---|---|---|
| $B_{ij}$ (GPa) | 141.3 | 159.28 | 70.97 | 62.94 | 32.14 |
| $dB_{ij}/dp$ | 2.86 | 3.04 | 2.44 | 2.86 | -0.22 |
| $d^2B_{ij}/dp^2$ | 0.14 | 0.176 | 0.12 | 0.12 | 0 |

Table S2: Effective elastic constants and their pressure derivatives for $\omega$-Zr that satisfy the consistency conditions.

| $\omega$-Zr | $B_{11}^e$ | $B_{33}^e$ | $B_{13}^e$ |
|---|---|---|---|
| $B_{ij}^e$ (GPa) | 169.95 | 168 | 72.67 |
| $dB_{ij}^e/dp$ | 1.84 | 2.02 | 1.47 |
| $d^2B_{ij}^e/dp^2$ | 0.06 | 0.076 | 0.068 |

## 2.2 Approximate Analytical Solution of Axisymmetric Problem on Compression of a Sample

We consider a polycrystalline material compressed by diamond anvils in axisymmetric formulation. We assume that macroscopically material behaves like perfectly plastic and isotropic. Such behavior can be achieved after large-enough preliminary plastic deformation [13, 14].

*System of equations and assumptions.* The pressure-dependent von Mises yield condition (i.e., Drucker-Prager yield condition) is assumed

$$(\sigma_{11} - \sigma_{22})^2 + (\sigma_{11} - \sigma_{33})^2 + (\sigma_{22} - \sigma_{33})^2 + 6\tau_{13}^2 = 2\sigma_y^2(p) = 6\tau_y^2(p), \tag{S.15}$$

where $\sigma_{33}$, $\sigma_{11}$ and $\sigma_{22}$ are the normal stress components along the load (vertical), radial and azimuthal directions, respectively (Fig. S5), $\tau = \tau_{13}$ is the shear stress, and $\sigma_y$ and $\tau_y$ are the yield strengths in compression and shear, respectively. Based on experimental results for $\alpha$- and $\omega$-Zr,

$$\sigma_y = \sqrt{3}\tau_y = \sigma_y^0 + bp, \tag{S.16}$$

where constants $\sigma_y^0$ and $b$ are from Fig. S2. Equilibrium equations are

$$\frac{\partial \sigma_{11}}{\partial r} + \frac{\partial \tau_{13}}{\partial z} + \frac{\sigma_{11} - \sigma_{22}}{r} = 0 \quad \text{in radial direction;} \tag{S.17}$$

$$\frac{\partial \sigma_{33}}{\partial z} + \frac{\partial \tau_{13}}{\partial r} + \frac{\tau_{13}}{r} = 0 \quad \text{in axial direction.} \tag{S.18}$$

The following assumptions are accepted:

1. As it approximately follows from DAC experiments and FEM simulations for polycrystals (based on the phenomenological flow theory of plasticity)

$$\sigma_{11} = \sigma_{22}. \tag{S.19}$$

Then the yield condition (S.15) simplifies to

$$(\sigma_{11} - \sigma_{33})^2 + 3\tau_{13}^2 = \sigma_y^2(p) = 3\tau_y^2(p). \tag{S.20}$$

2. Stress $\sigma_{33}$ is independent of $z$. This does not mean that

$$\frac{\partial \tau_{13}}{\partial r} + \frac{\tau_{13}}{r} = 0 \quad \rightarrow \quad \tau_{13} = \tau_0(z)\frac{r_0}{r}, \tag{S.21}$$

$r_0$ and $\tau_0(z)$ being arbitrary constants, because for material with pressure-independent yield strength, $\tau_0(z)$ at the contact surface $z = h$ ($h = h(r)$ is the half of the sample thickness profile determined from the experiments) may be equal to the constant $\tau_y$ for all $r$ at the contact surface. Approximate independence of $\sigma_{33}$ of $z$ means that two other terms in Eq. (S.18) make small contribution to $\sigma_{33}$.

*Solution.* A slightly modified Prandtl's solution (which was for a plane strain problem [15]) for stresses that satisfies equilibrium equations, plasticity condition, and the above assumptions is:

$$\frac{\sigma_{33}}{\tau_y} = \frac{\sigma_{33}^s}{\tau_y} + \frac{mr}{h}; \tag{S.22}$$

$$\frac{\tau_{13}}{\tau_y} = \frac{mz}{h}; \tag{S.23}$$

$$\frac{\sigma_{11}}{\tau_y} = \frac{\sigma_{33}^s}{\tau_y} + \frac{mr}{h} + \sqrt{3}\sqrt{1 - \left(\frac{mz}{h}\right)^2} = \frac{\sigma_{33}}{\tau_y} + \sqrt{3}\sqrt{1 - \left(\frac{mz}{h}\right)^2}; \tag{S.24}$$

$$p = -(2\sigma_{11} + \sigma_{33})/3, \tag{S.25}$$

where $\sigma_{33}^s$ is the stress $\sigma_{33}$ at the symmetry axis $r = 0$ and parameter $0 \leq m(r) \leq 1$ is defined by the value of shear stress $\tau_c$ at the contact surface, $\tau_c = m\tau_y$; $m(0) = 0$ at the symmetry axis. The difference with the Prandtl's solution is in multiplier $\sqrt{3}$ instead of 2 in Eq. (S.24) for $\sigma_{11}$. The reason is that we use von Mises condition and $\sigma_{11} = \sigma_{22}$, which results in Eq. (S.20), while in the Prandtl's solution the Tresca condition along with plane strain assumption lead to the yield condition $(\sigma_{11} - \sigma_{33})^2 + 4\tau_{13}^2 = \sigma_y^2 = 4\tau_y^2$.

*Averaging over the sample thickness.* For averaging all stresses over the sample thickness, to get transparent analytical results, we have to assume that the yield strength $\tau_y$ depends on the pressure averaged over the sample thickness,

$$\bar{p} = \frac{1}{h}\int_0^h p\,dz, \qquad (S.26)$$

i.e.,

$$\sigma_y = \sqrt{3}\tau_y = \sigma_y^0 + b\bar{p}. \qquad (S.27)$$

Then the stress $\bar{\sigma}_{11}$ averaged over the sample thickness is

$$\bar{\sigma}_{11} = \frac{1}{h}\int_0^h \sigma_{11}dz = \sigma_{33} + (\sigma_y^0 + b\bar{p})\frac{m\sqrt{1-m^2} + \arcsin(m)}{2m}. \qquad (S.28)$$

Averaged pressure

$$\bar{p} = -\frac{2}{3}\bar{\sigma}_{11} - \frac{1}{3}\sigma_{33} = -\sigma_{33} - \frac{2}{3}(\sigma_y^0 + b\bar{p})\frac{m\sqrt{1-m^2} + \arcsin(m)}{2m}, \qquad (S.29)$$

where we substituted Eq. (S.28) for $\bar{\sigma}_{11}$. Resolving this equation for $\sigma_{33}$, we obtain

$$\sigma_{33} = -\bar{p} - \frac{2}{3}(\sigma_y^0 + b\bar{p})\frac{m\sqrt{1-m^2} + \arcsin(m)}{2m}. \qquad (S.30)$$

Substituting $\sigma_{33}$ from Eq. (S.30) in Eq. (S.28) and resolving for $\bar{\sigma}_{11}$, we derive

$$\bar{\sigma}_{11} = -\bar{p} + \frac{1}{3}(\sigma_y^0 + b\bar{p})\frac{m\sqrt{1-m^2} + \arcsin(m)}{2m}. \qquad (S.31)$$

Thus, if friction stress in terms of $m$ and homogeneous along $z$ axis stress $\sigma_{33}$ are known, $\bar{p}$ and $\bar{\sigma}_{11}$ can be calculated.

Let us analyze the $z$-dependent part of normal stress and its averaged value:

$$\frac{\sigma_{11} - \sigma_{33}}{\sqrt{3}\tau_y} = \sqrt{1 - \left(\frac{mz}{h}\right)^2}; \qquad \frac{\bar{\sigma}_{11} - \sigma_{33}}{\sqrt{3}\tau_y} = \frac{m\sqrt{1-m^2} + \arcsin(m)}{2m}. \qquad (S.32)$$

Note that the averaged value $\bar{\sigma}_{11} - \sigma_{33}$ is much closer to the value of $\sigma_{11} - \sigma_{33}$ at the symmetry plane than at the contact surface. Indeed, at the symmetry plane $(\sigma_{11}(0) - \sigma_{33})/(\sqrt{3}\tau_y) = 1$ for all $m$; for example, for $m = 1$, we have at the contact surface $(\sigma_{11}(1) - \sigma_{33})/(\sqrt{3}\tau_y) = 0$, while averaged value is $(\bar{\sigma}_{11} - \sigma_{33})/(\sqrt{3}\tau_y) = 0.79$.

*Relation between averaged stresses and stresses at the contact surface and symmetry plane.* Our next objective is to find relationship between $\bar{\sigma}_{11}$ and $\sigma_{11}(0)$ and $\sigma_{11}(1)$ to be used for interpretation of experimental results. We will use the following identity

$$\bar{\sigma}_{11} = \sigma_{11}(1)w + \sigma_{11}(0)(1-w);$$
$$w(m) = \frac{\bar{\sigma}_{11} - \sigma_{11}(0)}{\sigma_{11}(1) - \sigma_{11}(0)} = \frac{\sqrt{1-m^2} + \arcsin(m)/m - 2}{2(\sqrt{1-m^2} - 1)}, \qquad (S.33)$$

where $w(m)$ is treated as the weight factor, which varies in a narrow range between $w(1) = 1 - \pi/4 \simeq 0.215$ and $w(0) = 1/3$. Since $\sigma_{33}$ is independent of $z$, we have similar equation for pressure:

$$\bar{p} = p(1)w + p(0)(1-w). \tag{S.34}$$

### 2.3 Special Stress States

Here we considered two main stress states under plastic deformation.

1. Along the symmetry axis ($r = m = 0$) and at the symmetry plane ($z = 0$), shear stress is zero, and the stress components along the load axis (Fig. S5) are defined by Eq. (S.24):

$$\sigma_{11}^s = \sigma_{33}^s + \sigma_y; \qquad \sigma_{11}(0) = \sigma_{22}(0) = \sigma_{33} + \sigma_y, \tag{S.35}$$

   where we took into account that $\sigma_{33}$ is assumed to be independent of $z$. Note that compressive normal stresses are negative. Then for pressure one obtains

$$p^s = p(0) = -(\sigma_{33} + 2/3\sigma_y(\bar{p})). \tag{S.36}$$

2. At sample-diamond contact surface $z/h = 1$, Eq. (S.24) results in

$$\sigma_{11}(1) = \sigma_{22}(1) = \sigma_{33} + \sigma_y(\bar{p})\sqrt{1-m^2}. \tag{S.37}$$

   and

$$p(1) = -(\sigma_{33} + 2/3\sigma_y(\bar{p})\sqrt{1-m^2}). \tag{S.38}$$

   When the friction shear stress reaches the yield strength in shear ($m = 1$), the von Mises yield condition (S.15) results in

$$\sigma_{11}(1) = \sigma_{22}(1) = \sigma_{33}(1) = -p(1) \tag{S.39}$$

   without the assumption $\sigma_{11} = \sigma_{22}$.

### 2.4 Application of the modified Hooke's law for $\alpha$-Zr with $c$-axis parallel to the loading axis

The modified Prandtl's solution in Section 2.2 allows us to find stress distributions provided that the boundary conditions $m(r)$ for the contact friction and $\sigma_{33}^s$ are known, but they are not. We have to find a way to utilize experimentally measured fields $\overline{E}_{0,rr}$ and $\overline{E}_{0,\theta\theta}$ to determine $m(r)$ and finalize our analytical solution. Strong observed texture (with $c$-axis parallel to the loading axis for $\alpha$-Zr and to the radial direction for $\omega$-Zr) allows us to utilize single crystal elasticity to find stress distributions in a sample using modified Hooke's law at high pressure. Then the Reuss hypothesis that stresses in all single crystals in the representative volume and in polycrystalline aggregate are the same, combined with the simplified mechanical equilibrium condition (S.59) allow us to connect these stresses and stresses determined by the modified Prandtl's solution and determine $m(r)$, and consequently, all stress and elastic strain fields.

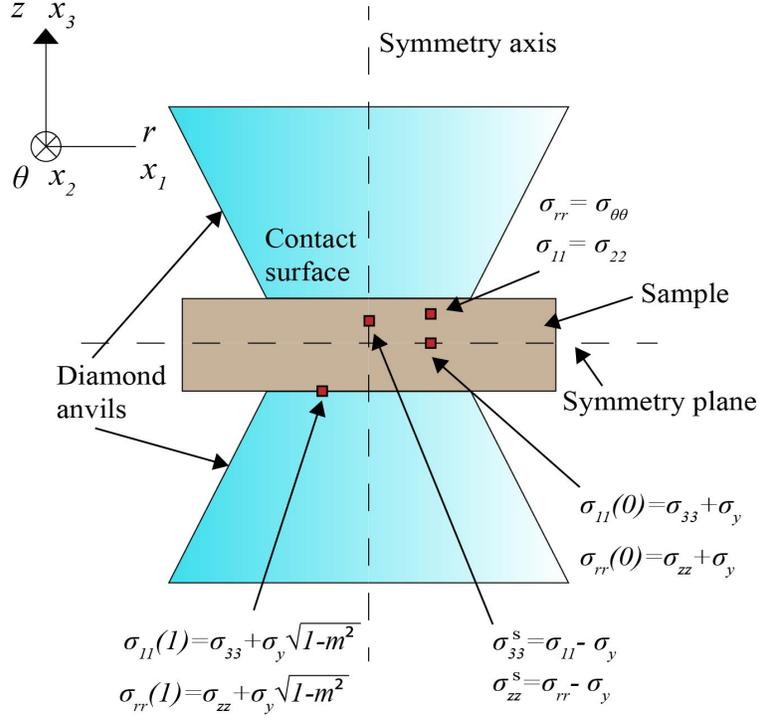

Figure S5: Schematic illustration of stress states at the symmetry plane, symmetry axis, and the contact surface.

The Hooke's law (S.14) for hexagonal crystal, in a hydrostatically stressed configuration characterized by pressure $p$, for normal stresses and strains can be presented in the form

$$\begin{bmatrix} \sigma_{11} + p \\ \sigma_{22} + p \\ \sigma_{33} + p \end{bmatrix} = \begin{bmatrix} \hat{\sigma}_{11} \\ \hat{\sigma}_{22} \\ \hat{\sigma}_{33} \end{bmatrix} = \begin{bmatrix} B_{11} & B_{12} & B_{13} \\ B_{12} & B_{11} & B_{13} \\ B_{13} & B_{13} & B_{33} \end{bmatrix} \begin{bmatrix} \tilde{\varepsilon}_{11} \\ \tilde{\varepsilon}_{22} \\ \tilde{\varepsilon}_{33} \end{bmatrix}. \tag{S.40}$$

We will not use the second equation (S.14) for shear strains and stresses because shear strains are not measurable with x-rays and cannot be used for connection to the Prandtl's solution. However, after all stresses, including shear stress will be found, this equation can be used to determine the elastic shear strain.

Using our assumption (S.19) that $\sigma_{11} = \sigma_{22}$ for a polycrystal and by invoking the Reuss hypothesis, $\sigma_{11} = \sigma_{22}$ in the single crystal is obtained. This equality, using Eq. (S.40), gives

$$\tilde{\varepsilon}_{11} = \tilde{\varepsilon}_{22} \quad \rightarrow \quad \bar{\tilde{\varepsilon}}_{11} = \bar{\tilde{\varepsilon}}_{22}. \tag{S.41}$$

From Eq. (S.13) and $F_{*11} = F_{*22}$ we obtain $F_{0,11} = F_{0,22}$ and $E_{0,11} = E_{0,22}$. Averaging these equations over a sample thickness leads to the experimental observation that $\overline{F}_{0,11} = \overline{F}_{0,22}$ and $\overline{E}_{0,11} = \overline{E}_{0,22}$.

For $\sigma_{11} = \sigma_{22}$, one obtains from Eq. (S.40)

$$\tilde{\varepsilon}_{11}(z/h) = \tilde{\varepsilon}_{22}(z/h) = k(B_{33}\hat{\sigma}_{11}(z/h) - B_{13}\hat{\sigma}_{33});$$
$$\tilde{\varepsilon}_{33}(z/h) = k((B_{11}+B_{12})\hat{\sigma}_{33} - 2B_{13}\hat{\sigma}_{11}(z/h)); \quad k = \left((B_{11}+B_{12})B_{33} - 2B_{13}^2\right)^{-1} \quad \text{(S.42)}$$

*Averaging over the sample thickness.* Averaging Eq. (S.42) over the thickness, we derive

$$\bar{\tilde{\varepsilon}}_{11} = k(B_{33}(\bar{\sigma}_{11} + \bar{p}) - B_{13}(\bar{\sigma}_{33} + \bar{p})); \quad \text{(S.43)}$$
$$\bar{\tilde{\varepsilon}}_{33} = k((B_{11}+B_{22})(\bar{\sigma}_{33}+\bar{p}) - 2B_{13}(\bar{\sigma}_{11}+\bar{p})). \quad \text{(S.44)}$$

Note that elastic moduli $\boldsymbol{B}$ are considered homogeneous along the thickness direction for simplicity; they are also considered to be functions of averaged through the thickness pressure $\bar{p}$, i.e., $\boldsymbol{B} = \boldsymbol{B}(\bar{p})$, where $\bar{p} = \bar{p}(r)$. To prove for elastic strains relationship similar to Eq. (S.33) for stresses, let us derive an alternative expression for $\bar{\tilde{\varepsilon}}_{11}$. It follows from Eq. (S.42)

$$\tilde{\varepsilon}_{11}(0) = k(B_{33}\hat{\sigma}_{11}(0) - B_{13}\hat{\sigma}_{33}(0)) = k(B_{33}(\sigma_{11}(0) + p(0)) - B_{13}(\sigma_{33} + p(0))); \quad \text{(S.45)}$$
$$\tilde{\varepsilon}_{11}(1) = k(B_{33}(\sigma_{11}(1) + p(1)) - B_{13}(\sigma_{33} + p(1))). \quad \text{(S.46)}$$

We remind that the arguments 0 and 1 designate points at the symmetry plane and the contact surface, respectively. Multiplying Eq. (S.45) by $1 - w$ and Eq. (S.46) by $w$ and adding them, we obtain

$$(1-w)\tilde{\varepsilon}_{11}(0) + w\tilde{\varepsilon}_{11}(1) = k[B_{33}\left((1-w)\sigma_{11}(0) + w\sigma_{11}(1)\right) + B_{33}\left((1-w)p(0) + wp(1)\right) -$$
$$B_{13}\left((1-w)\sigma_{33} + w\sigma_{33} + (1-w)p(0) + wp(1)\right)]. \quad \text{(S.47)}$$

Using Eq. (S.34) in Eq. (S.47), we obtain

$$(1-w)\tilde{\varepsilon}_{11}(0) + w\tilde{\varepsilon}_{11}(1) = k[B_{33}(\bar{\sigma}_{11} + \bar{p}) - B_{13}(\sigma_{33} + \bar{p})], \quad \text{(S.48)}$$

where the first equation Eq. (S.33) was utilized. Comparison of Eq. (S.48) and Eq. (S.43) leads to important conclusion that the averaged small imposed strains,

$$\bar{\tilde{\varepsilon}}_{11} = \bar{\tilde{\varepsilon}}_{22} = (1-w)\tilde{\varepsilon}_{11}(0) + w\tilde{\varepsilon}_{11}(1), \quad \text{(S.49)}$$

have similar expressions as the averaged stress $\bar{\sigma}_{11}$ in Eq. (S.33).This result will be utilized for interpretation of experimental measurements of $\bar{\tilde{\varepsilon}}_{11}$. Similarly, for $\bar{\tilde{\varepsilon}}_{33}$

$$\bar{\tilde{\varepsilon}}_{33} = (1-w)\tilde{\varepsilon}_{33}(0) + w\tilde{\varepsilon}_{33}(1). \quad \text{(S.50)}$$

*At the symmetry plane*, using $\tilde{\varepsilon}_{11}(0) = \tilde{\varepsilon}_{22}(0)$, $\sigma_{11}(0) = \sigma_{22}(0)$ and Eq. (S.35) in Eq. (S.40), we derive

$$\begin{bmatrix} \sigma_{33} + \sigma_y(\bar{p}) + p(0) \\ \sigma_{33} + \sigma_y(\bar{p}) + p(0) \\ \sigma_{33} + p(0) \end{bmatrix} = \begin{bmatrix} B_{11} & B_{12} & B_{13} \\ B_{12} & B_{11} & B_{13} \\ B_{13} & B_{13} & B_{33} \end{bmatrix} \begin{bmatrix} \tilde{\varepsilon}_{11}(0) \\ \tilde{\varepsilon}_{11}(0) \\ \tilde{\varepsilon}_{33}(0) \end{bmatrix}. \quad \text{(S.51)}$$

*At the sample-anvil contact surface*, using Eqs. (S.37), we obtain from Eqs. (S.40)

$$\begin{bmatrix} \sigma_{33} + \sigma_y(\bar{p})\sqrt{1-m^2} + p(1) \\ \sigma_{33} + \sigma_y(\bar{p})\sqrt{1-m^2} + p(1) \\ \sigma_{33} + p(1) \end{bmatrix} = \begin{bmatrix} B_{11} & B_{12} & B_{13} \\ B_{12} & B_{11} & B_{13} \\ B_{13} & B_{13} & B_{33} \end{bmatrix} \begin{bmatrix} \tilde{\varepsilon}_{11}(1) \\ \tilde{\varepsilon}_{11}(1) \\ \tilde{\varepsilon}_{33}(1) \end{bmatrix}. \tag{S.52}$$

In order to utilize averaged over the thickness elastic strains $\bar{\tilde{\varepsilon}}_{11} = \bar{\tilde{\varepsilon}}_{22}$ (which are related to $\overline{E}_{0,11} = \overline{E}_{0,22}$ and $\overline{F}_{*11} = \overline{F}_{*22}$ measured in experiments through Eq. (S.10)), let us multiply Eq. (S.51) by $1-w$, Eq. (S.52) by $w$, and combine them:

$$\begin{bmatrix} B_{11} & B_{12} & B_{13} \\ B_{12} & B_{11} & B_{13} \\ B_{13} & B_{13} & B_{33} \end{bmatrix} \begin{bmatrix} \bar{\tilde{\varepsilon}}_{11} \\ \bar{\tilde{\varepsilon}}_{11} \\ \bar{\tilde{\varepsilon}}_{33} \end{bmatrix} = \begin{bmatrix} \sigma_{33} + (1-w)\sigma_y(\bar{p}) + w\sigma_y(\bar{p})\sqrt{1-m^2} + (1-w)p(0) + wp(1) \\ \sigma_{33} + (1-w)\sigma_y(\bar{p})) + w\sigma_y(\bar{p})\sqrt{1-m^2} + (1-w)p(0) + wp(1) \\ \sigma_{33} + (1-w)p(0) + wp(1) \end{bmatrix}$$

$$= \begin{bmatrix} \sigma_{33} + (1-w)\sigma_y(\bar{p}) + w\sigma_y(\bar{p})\sqrt{1-m^2} + \bar{p} \\ \sigma_{33} + (1-w)\sigma_y(\bar{p}) + w\sigma_y(\bar{p})\sqrt{1-m^2} + \bar{p} \\ \sigma_{33} + \bar{p} \end{bmatrix}. \tag{S.53}$$

In obtaining Eq. (S.53), Eqs. (S.34), (S.49) and (S.50) were used. The advantage of Eq. (S.53) is that it includes $\bar{\tilde{\varepsilon}}_{11}$, which will allow us to express all stresses and strains in terms of $\bar{\tilde{\varepsilon}}_{11}$. Inversion of Eqs. (S.53) gives:

$$\bar{\tilde{\varepsilon}}_{33} = k(B_{11} + B_{12} - 2B_{13})(\sigma_{33} + \bar{p}) - 2kB_{13}\left[(1-w)\sigma_y(\bar{p}) + w\sigma_y(\bar{p})\sqrt{1-m^2}\right]. \tag{S.54}$$

Also, the third Eq. (S.53) is

$$\sigma_{33} + \bar{p} = 2B_{13}\bar{\tilde{\varepsilon}}_{11} + B_{33}\bar{\tilde{\varepsilon}}_{33}. \tag{S.55}$$

Substituting Eq. (S.55) in Eq. (S.54) and solving for $\bar{\tilde{\varepsilon}}_{33}$, one finds

$$\bar{\tilde{\varepsilon}}_{33} = (B_{33} - B_{13})^{-1}\left[(B_{11} + B_{12} - 2B_{13})\bar{\tilde{\varepsilon}}_{11} - (1-w)\sigma_y(\bar{p}) - w\sigma_y(\bar{p})\sqrt{1-m^2}\right]. \tag{S.56}$$

Placing Eq. (S.56) in Eq. (S.55), we obtain expression for $\sigma_{33} + \bar{p}$:

$$\sigma_{33} + \bar{p} = \tag{S.57}$$
$$(B_{13} - B_{33})^{-1}\left[(2B_{13}^2 - B_{11}B_{33} - B_{12}B_{33})\bar{\tilde{\varepsilon}}_{11} + B_{33}\sigma_y(\bar{p})\left(1 - w + w\sqrt{1-m^2}\right)\right].$$

Next, we substitute $\sigma_{33} + \bar{p}$ from Eq. (S.30) in Eq. (S.57) and obtain nonlinear equation for $\bar{p}$ (if we assume $m$ to be known):

$$(B_{13} - B_{33})^{-1}\left[(2B_{13}^2 - B_{11}B_{33} - B_{12}B_{33})\bar{\tilde{\varepsilon}}_{11} + B_{33}\sigma_y(\bar{p})\left(1 - w + w\sqrt{1-m^2}\right)\right]$$
$$= -\frac{2}{3}(\sigma_y^0 + b\bar{p})\frac{m\sqrt{1-m^2} + \arcsin(m)}{2m}. \tag{S.58}$$

Note that $B_{ij}$ are quadratic functions of pressure, and due to nonlinearity of Eq. (S.58) in $B_{ij}$, it is strongly nonlinear in $\bar{p}$ and cannot be solved analytically. After numerical solution of Eq. (S.58) for $\bar{p}$, Eq. (S.30) gives numerical value of $\sigma_{33}$. Note that since experimentally determined strains $\bar{\varepsilon}_{11} = \bar{\varepsilon}_{22}$ are functions of the radius $r$, Eq. (S.57) and solution to Eq. (S.58) also reproduce the radial dependence of $\sigma_{33}$ and $\bar{p}$. Then Eqs. (S.23), (S.24), and (S.25) can be used to reproduce $z$-dependence of the shear stress $\tau_{13}$, normal stresses $\sigma_{11} = \sigma_{22}$, and pressure $p$. Elastic strains can be found after substituting stresses in Eqs. (S.42). The next step is to derive the additional equation for $m$ to be combined with Eq. (S.58).

## 2.5 Evaluation of friction stress utilizing pressure gradient method

The friction stress $\tau_c$ is defined by the simplified mechanical equilibrium equation [13]

$$\frac{d\bar{\sigma}_{11}}{dr} = \frac{\tau_c}{h(r)} = \frac{m(r)\tau_y(\bar{p})}{h(r)}. \tag{S.59}$$

For solution, we approximated this differential equation with the finite difference equation, using midpoint algorithm, e.g.,

$$\frac{\bar{\sigma}_{11,i+1} - \bar{\sigma}_{11,i-1}}{2\Delta r} = \frac{m_i(\sigma_y^0 + b\bar{p}_i)}{\sqrt{3}h_i}, \tag{S.60}$$

where subscript $i$ designates value of the function at point $r_i$. This equation is supplemented with Eqs. (S.31) and (S.58) for each radial point $r_i$:

$$\bar{\sigma}_{11} = -\bar{p} + \frac{1}{3}(\sigma_y^0 + b\bar{p})\frac{m\sqrt{1-m^2} + \arcsin(m)}{2m}. \tag{S.61}$$

$$(B_{13} - B_{33})^{-1}\left[(2B_{13}^2 - B_{11}B_{33} - B_{12}B_{33})\bar{\bar{\varepsilon}}_{11} + B_{33}\sigma_y(\bar{p})\left(1 - w + w\sqrt{1-m^2}\right)\right]$$
$$= -\frac{2}{3}(\sigma_y^0 + b\bar{p})\frac{m\sqrt{1-m^2} + \arcsin(m)}{2m}. \tag{S.62}$$

Eqs. (S.60)-(S.62) represent system of 3N nonlinear algebraic/trigonometric equations for 3 unknowns $m$, $\bar{p}$, and $\bar{\sigma}_{11}$ in each of N points along the radial direction $r$, which we solved numerically. Since for $\bar{\sigma}_{11}$ we have explicit expression, which in practice is substituted in Eq. (S.60), we have 2N nonlinear algebraic/trigonometric equations for 2 unknowns $m$ and $\bar{p}$ in each of N points. Then Eqs. (S.23), (S.24), and (S.25) can be used to reproduce $z$-dependence of the shear stress $\tau_{13}$, normal stresses $\sigma_{11} = \sigma_{22}$, and pressure $p$. Elastic strains can be found after substituting stresses in Eqs. (S.42).

We can pass to FEM simulations, as one of the boundary conditions, either obtained contact shear stress $\tau_c(r)$ or coefficient $m(r)$. In the latter case, it is more precise to redefine $m$ based on the local pressure $p(1)$, i.e., from $\tau_c(r) = m(r)\tau_y(\bar{p}) = m'(r)\tau_y[p(1)]$ and use in FEM simulations $m'$, because in FEM $\tau_y$ at the contact surface depends on $p(1)$. Pressure $p(1)$ can be obtained from Eq. (S.38), in which $\sigma_{33}$ is substituted with the expression

(S.30)):

$$p(1) = \bar{p} - \frac{1}{3}\sigma_y(\bar{p})\frac{m\sqrt{1-m^2} - \arcsin(m)}{m}. \tag{S.63}$$

## 2.6 Application of the Hooke's law for $\omega$-Zr with $c$-axis parallel to the radial direction

The modified Hooke's law for hexagonal crystal for normal stresses and strains, when $c$-axis is parallel to the radial direction 1 can be presented in the form

$$\begin{bmatrix} \sigma_{11} + p \\ \sigma_{22} + p \\ \sigma_{33} + p \end{bmatrix} = \begin{bmatrix} \hat{\sigma}_{11} \\ \hat{\sigma}_{22} \\ \hat{\sigma}_{33} \end{bmatrix} = \begin{bmatrix} B_{33} & B_{13} & B_{13} \\ B_{13} & B_{11} & B_{12} \\ B_{13} & B_{12} & B_{11} \end{bmatrix} \begin{bmatrix} \tilde{\varepsilon}_{11} \\ \tilde{\varepsilon}_{22} \\ \tilde{\varepsilon}_{33} \end{bmatrix}. \tag{S.64}$$

Our assumption (S.19) $\sigma_{11} = \sigma_{22}$ for a polycrystal along with the Reuss hypothesis leads to $\sigma_{11} = \sigma_{22}$ for the single crystal. The problem is that this equality substituted in Eq. (S.64) does not lead to $\tilde{\varepsilon}_{11} = \tilde{\varepsilon}_{22}$ and hence, violates the experimental observation that $\overline{E}_{0,11} = \overline{E}_{0,22}$. That is we need to modify the elastic moduli tensor $\boldsymbol{B}$ for consistency with experiments.

The simplest way to satisfy $\sigma_{11} = \sigma_{22}$ for polycrystal is to assume that for $\tilde{\varepsilon}_{11} = \tilde{\varepsilon}_{22}$ stresses $\sigma_{11}^p$ and $\sigma_{22}^p$ for polycrystal are defined as

$$\sigma_{11}^p = \sigma_{22}^p = 0.5(\sigma_{11} + \sigma_{22}) \quad for \quad \tilde{\varepsilon}_{11} = \tilde{\varepsilon}_{22}. \tag{S.65}$$

This can be imposed by changing the elastic moduli matrix in Eq. (S.64) with an "effective" elastic modular matrix

$$\begin{bmatrix} \sigma_{11} + p \\ \sigma_{22} + p \\ \sigma_{33} + p \end{bmatrix} = \begin{bmatrix} \hat{\sigma}_{11} \\ \hat{\sigma}_{22} \\ \hat{\sigma}_{33} \end{bmatrix} = \begin{bmatrix} B_{11}^e & B_{12}^e & B_{13}^e \\ B_{12}^e & B_{11}^e & B_{13}^e \\ B_{13}^e & B_{13}^e & B_{33}^e \end{bmatrix} \begin{bmatrix} \tilde{\varepsilon}_{11} \\ \tilde{\varepsilon}_{22} \\ \tilde{\varepsilon}_{33} \end{bmatrix} = \tag{S.66}$$

$$\begin{bmatrix} 0.5(B_{11}+B_{33}) & B_{13} & 0.5(B_{12}+B_{13}) \\ B_{13} & 0.5(B_{11}+B_{33}) & 0.5(B_{12}+B_{13}) \\ 0.5(B_{12}+B_{13}) & 0.5(B_{12}+B_{13}) & B_{11} \end{bmatrix} \begin{bmatrix} \tilde{\varepsilon}_{11} \\ \tilde{\varepsilon}_{22} \\ \tilde{\varepsilon}_{33} \end{bmatrix}.$$

Thus, we substitute elastic constants in positions 11 and 22 in matrix (S.64) (i.e., $B_{33}$ and $B_{11}$) with their average $0.5(B_{11}+B_{33})$, and elastic constants in positions 13 and 23 in matrix (S.64) (i.e., $B_{13}$ and $B_{12}$) with their average $0.5(B_{12}+B_{13})$, keeping symmetry of the elasticity matrix. Matrix (S.66) has 4 independent elastic constants, like matrix (S.64). However, with such a procedure, we violated equality of elastic moduli in positions 12 and 13 in Eq. (S.64).

To restore this equality (symmetry), we accept

$$\begin{bmatrix}\sigma_{11}+p\\ \sigma_{22}+p\\ \sigma_{33}+p\end{bmatrix}=\begin{bmatrix}B^e_{11} & B^e_{13} & B^e_{13}\\ B^e_{13} & B^e_{11} & B^e_{13}\\ B^e_{13} & B^e_{13} & B^e_{33}\end{bmatrix}\begin{bmatrix}\tilde{\varepsilon}_{11}\\ \tilde{\varepsilon}_{22}\\ \tilde{\varepsilon}_{33}\end{bmatrix}= \qquad (S.67)$$

$$\begin{bmatrix}0.5(B_{11}+B_{33}) & 0.5(B_{12}+B_{13}) & 0.5(B_{12}+B_{13})\\ 0.5(B_{12}+B_{13}) & 0.5(B_{11}+B_{33}) & 0.5(B_{12}+B_{13})\\ 0.5(B_{12}+B_{13}) & 0.5(B_{12}+B_{13}) & B_{11}\end{bmatrix}\begin{bmatrix}\tilde{\varepsilon}_{11}\\ \tilde{\varepsilon}_{22}\\ \tilde{\varepsilon}_{33}\end{bmatrix}.$$

Reduction in number of independent elastic moduli for normal strains from 4 to 3 in transition from a single crystal to textured polycrystal is natural. After such a modification of the $\boldsymbol{B}$-matrix, all equations (S.42)-(S.62) for $\alpha$-Zr can be applied with adding superscript $e$ to components $B_{ij}$ and imposing $B_{12}=B_{13}$.

### 2.7 Generalization for the two-phase mixture

*Equations for each phase.* Radial distribution of volume fraction of the low- and high-pressure phases, averaged over the sample thickness, $c_1$ and $c_2=1-c_1=c$, are determined from the experiment using Rietveld refinement. It is independent of the interpretation of stresses and strains and does not participate in the iterations to determine distribution of the stress and elastic strain tensors. Radial distributions of strain $\bar{E}^1_{0,11}=\bar{E}^2_{022}$ is measured in each phase 1 and 2. The modified Hooke's law and plasticity condition are satisfied for each phase separately. We assume

$$m_1=m_2=m \qquad (S.68)$$

and that $\sigma^k_{11}=\sigma^k_{22}$, $\sigma^k_{33}$ is independent of $z$, and Eqs. (S.22) - (S.24) are valid for stresses in each phase. Then Eqs. (S.22) - (S.67) are valid for each phase with the elastic constants and the yield strength of each phase, all in terms of experimentally measured elastic strains $\bar{E}^k_{0,11}=\bar{E}^k_{0,22}$ in each phase. That means that for two-phase material, we have to perform the same procedure and use the same equations for each phase separately and find solution in each phase separately. This does not mean that we completely neglect interaction between phases, because the experimentally measured elastic strains $\bar{E}^k_{0,11}=\bar{E}^k_{0,22}$ in each phase do include such an interaction.

*Equations for mixture.* Since all fields in each phase are known, here we define averaging rules to determine fields in the mixture. We assume for deformation gradient under hydrostatic pressure $\boldsymbol{F}_*$, elastic superposed strains $\varepsilon_{ii}$, and for stresses $\sigma_{ii}$

$$\boldsymbol{F}_*=c_1\boldsymbol{F}^1_*+c_2\boldsymbol{F}^2_*; \qquad E_{0,ii}=c_1E^1_{0,ii}+c_2E^2_{0,ii}; \qquad \gamma=c_1\gamma^1+c_2\gamma^2;$$
$$\sigma_{ii}=c_1\sigma^1_{ii}+c_2\sigma^2_{ii}; \qquad \tau=c_1\tau^1+c_2\tau^2; \qquad \tilde{\varepsilon}_{ii}=c_1\tilde{\varepsilon}^1_{ii}+c_2\tilde{\varepsilon}^2_{ii}. \qquad (S.69)$$

While for stresses the averaging equation is exact, for elastic strain and $\boldsymbol{F}_*$ they represent strong assumption, because $\boldsymbol{F}_*$ and $\tilde{\boldsymbol{\varepsilon}}$ are incompatible separately even in the absence of the plastic and transformation strains, because of heterogeneous pressure distribution, and consequently $\boldsymbol{F}_*$. For shear stress at the contact surface, in particular,

$$\tau_c = m\tau_y = c_1\tau_{c1}(\bar{p}_1) + c_2\tau_{c2}(\bar{p}_2) = m(c_1\tau_{y1}(\bar{p}_1) + c_2\tau_{y2}(\bar{p}_2)), \tag{S.70}$$

where from we obtain the mixture rule for the pressure-dependent yield strength

$$\tau_y = c_1\tau_{y1}(\bar{p}_1) + c_2\tau_{y2}(\bar{p}_2);$$
$$\sigma_y = \sqrt{3}\tau_y = c_1\sigma_{y1}(\bar{p}_1) + c_2\sigma_{y2}(\bar{p}_2) = c_1\sigma_{y1}^0 + c_2\sigma_{y2}^0 + c_1b_1\bar{p}_1 + c_2b_2\bar{p}_2. \tag{S.71}$$

## 2.8 Explicit relationships between $\tilde{\varepsilon}$ and $E$ and $\boldsymbol{F}_*$

It follows from Eq. (S.10) or (S.13) that

$$\tilde{\varepsilon}_{11} = \tilde{\varepsilon}_{22} = \frac{E_{0,11} - 0.5\left((F_{*11}(p))^2 - 1\right)}{(F_{*11}(p))^2}, \tag{S.72}$$

where for $\alpha - Zr$ $F_{*11}(p)$ is given by Eq. (S.4). It is important to note that the $\alpha - Zr$ crystal coordinate system coincides with the global coordinate system and therefore the local $F_{*11}(p)$ coincides with the global in Eq. (S.72). Averaged over the sample thickness $\bar{\tilde{\varepsilon}}_{11}$, which is present in Eq. (S.56), can be obtained from Eq. (S.72) as follows:

$$\bar{\tilde{\varepsilon}}_{11} = \frac{1}{h}\int_0^h \left(\frac{E_{0,11}}{(F_{*11}(p))^2} + \frac{1}{2}\frac{1}{(F_{*11}(p))^2} - \frac{1}{2}\right)dz \tag{S.73}$$

Based on Eq. (S.4), we can approximate in the pressure range of $0 - 10$ GPa as $\frac{1}{(F_{*11}(p))^2} = 1 + 0.00668p$. Using this, Eq. (S.73) can be written as

$$\bar{\tilde{\varepsilon}}_{11} = \frac{0.00668}{h}\int_0^h E_{0,11}p\,dz + \overline{E}_{0,11} + 0.00334\bar{p}. \tag{S.74}$$

It is impossible to evaluate $\int_0^h E_{0,11}p\,dz$ in terms of $\bar{p}$ and $\overline{E}_{0,11}$ without knowing the distribution of $E_{0,11}$ and therefore, it is approximated as:

$$\frac{1}{h}\int_0^h E_{0,11}p\,dz \simeq \overline{E}_{0,11}\bar{p}. \tag{S.75}$$

Combination of Eqs. (S.75) and (S.74) results in

$$\bar{\tilde{\varepsilon}}_{11} = 0.00668\overline{E}_{0,11}\bar{p} + \overline{E}_{0,11} + 0.00334\bar{p}. \tag{S.76}$$

As it was already mentioned that the *c*-axis of $\omega$-Zr crystal is primarily oriented along the global 1 direction and therefore, it is local $F_{*33}(p)$ (rather than the local $F_{*11}(p)$) that coincides with the global 1 direction, unlike for the $\alpha$-Zr. Also, to enforce $\tilde{\varepsilon}_{11} = \tilde{\varepsilon}_{22}$, given

$\overline{E}_{0,11} = \overline{E}_{0,22}$ from experiments, Eq. (S.72) transforms for $\omega$-Zr as:

$$\tilde{\varepsilon}_{11} = \tilde{\varepsilon}_{22} = \frac{E_{0,11} - 0.5\left(\left(F^{ef}_{*11}(p)\right)^2 - 1\right)}{\left(F^{ef}_{*11}(p)\right)^2}, \tag{S.77}$$

where, $F^{ef}_{*11} = F^{ef}_{*22} = 0.5(F^{\omega}_{*11} + F^{\omega}_{*33})$. $F^{\omega}_{*11}(p)$ and $F^{\omega}_{*33}(p)$ given in Eqs. (S.6) and (S.7) are written here again for convenience:

$$F^{\omega}_{*11}(p) = 1.00 - 0.00330p + 0.00003p^2, \tag{S.78}$$

and

$$F^{\omega}_{*33}(p) = 1.0 - 0.00282p + 0.00002p^2. \tag{S.79}$$

Following the same averaging procedure as was done with $\alpha$-Zr, for a pressure range of $0 - 20$ GPa, the approximation for $\omega$-Zr is

$$\frac{1}{\left(F^{ef}_{*11}(p)\right)^2} = 1.0 + 0.0055p \tag{S.80}$$

resulting in

$$\overline{\tilde{\varepsilon}}_{11} = \overline{\tilde{\varepsilon}}_{22} = 0.0055\overline{E}_{0,11}\overline{p} + \overline{E}_{0,11} + 0.0028\overline{p}. \tag{S.81}$$

Likewise, from Eq. (S.10) for direction 3 we have:

$$E_{0,33} = \tilde{\varepsilon}_{33}\left(F_{*33}(p)\right)^2 + 0.5((F_{*33}(p))^2 - 1). \tag{S.82}$$

Eq. (S.82) is not used in further derivations; it just defines the total Lagrangian strain $E_{0,33}$ when $\tilde{\varepsilon}_{33}$ is already found. Since function $F_{*33}$ is given by Eq. (S.5) for $\alpha - Zr$, it can be approximated by

$$(F_{*33}(p))^2 = 1 - 0.00664p + 0.00013p^2. \tag{S.83}$$

For the rotated $\omega - Zr$ crystal, the $E_{0,33}$ in the global coordinate system is as follows:

$$E_{0,33} = \tilde{\varepsilon}_{33}\left(F^{ef}_{*33}(p)\right)^2 + 0.5\left(\left(F^{ef}_{*33}(p)\right)^2 - 1\right), \tag{S.84}$$

where $F^{ef}_{*33} = F^{\omega}_{*11}$ and $\left(F^{ef}_{*33}(p)\right)^2 = 1 - 0.0066p + 0.000077p^2$ for the $\omega - Zr$.

For averaging over the thickness, we use $p$ from the Prandtl's solution Eq. (S.25):

$$p = -\frac{1}{3}(2\sigma_{11} + \sigma_{33}) = -\sigma_{33} - \frac{2}{3}\sigma_y\sqrt{1 - \left(\frac{mz}{h}\right)^2}. \tag{S.85}$$

Then
$$p^2 = \sigma_{33}^2 + \frac{4}{9}\sigma_y^2\left(1 - \left(\frac{mz}{h}\right)^2\right) + \frac{4}{3}\sigma_{33}\sigma_y\sqrt{1 - \left(\frac{mz}{h}\right)^2}. \quad (S.86)$$

After averaging over the thickness we obtain:
$$\overline{p^2} = \sigma_{33}^2 + \frac{4}{9}\sigma_y^2(\bar{p})\left(1 - \frac{m^2}{3}\right) + \frac{4}{3}\sigma_{33}\sigma_y(\bar{p})\frac{m\sqrt{1 - m^2} + sin^{-1}m}{2m}. \quad (S.87)$$

It is clear that $\overline{p^2} \neq \bar{p}^2$. Using Eq. (S.82), the average over the thickness total Lagrangian strain is
$$\overline{E}_{0,33} = \bar{\tilde{\varepsilon}}_{33} - 0.00664\bar{\tilde{\varepsilon}}_{33}\bar{p} + 0.00013\bar{\tilde{\varepsilon}}_{33}\overline{p^2} + 0.5(0.00013\overline{p^2} - 0.00664\bar{p}) \quad (S.88)$$

for $\alpha$-Zr and

$$\overline{E}_{0,33} = \bar{\tilde{\varepsilon}}_{33} - 0.0066\bar{\tilde{\varepsilon}}_{33}\bar{p} + 0.000077\bar{\tilde{\varepsilon}}_{33}\overline{p^2} + 0.5(0.000077\overline{p^2} - 0.0066\bar{p}) \quad (S.89)$$

for $\omega$-Zr. In obtaining Eqs. (S.88) and (S.89), assumptions $\frac{1}{h}\int_0^h \tilde{\varepsilon}_{33}p\,dz \simeq \bar{\tilde{\varepsilon}}_{33}\bar{p}$ and $\frac{1}{h}\int_0^h \tilde{\varepsilon}_{33}p^2\,dz \simeq \bar{\tilde{\varepsilon}}_{33}\overline{p^2}$ are used.

## 3 The reasons for difference between stresses and elastic strains from Rietveld refinement and CEA approach

To better illustrate the main sources and reasons for the difference between two approaches, let us calculate averaged over thickness stresses at the sample center, where shear strain, stresses, and $m$ are zero, see Fig. S6. Three approaches will be compared.

1. Using standard Rietveld refinement, the ratios of lattice parameters for the chosen sample thickness of $\alpha$-Zr are $a/a_0 = 0.996$, $c/c_0 = 0.995$, and for the corresponding unit cell volume ratio $V/V_0 = 0.988$. Utilization of the experimental generalized EOSs for $a$, $c$, and $V$ results in three different pressures, $p_a = 0.987$ GPa, $p_c = 1.463$ GPa, and $p_V = 1.138$ GPa, respectively. Difference between these values shows the inconsistency of utilizing EOS obtained under hydrostatic conditions to determine pressure under non-hydrostatic conditions, especially with the axial diffraction.

2. If we use $\bar{E}_{0,rr} = -0.0032$ and $\bar{E}_{0,\theta\theta} = -0.0033$ measured experimentally (which are consistent with $a/a_0 = 0.996$ and c-axis aligned with z-axis), and determine $\bar{E}_{0,zz} = -0.0055$ to have the same $V/V_0 = 0.988$ from the standard Rietveld refinement (since $\bar{E}_{0,zz}$ does not contribute to the X-ray diffraction patterns in the axial geometry), then applying the modified pressure-dependent Hooke's law (assuming $\bar{E}_{0,rr} \approx \bar{E}_{0,\theta\theta} = 0.5(\bar{E}_{0,rr} + \bar{E}_{0,\theta\theta}))$, we obtain $\sigma_{rr} = \sigma_{\theta\theta} = $ -1.07 GPa, $\sigma_{zz} = $-1.33 GPa, and $p=1.15$ GPa. Thus, $p$ did not practically change in comparison with $p_V$. The main problem is that $|\sigma_{zz} - \sigma_{rr}| = 0.26$ GPa, which is much smaller than the yield strength $\sigma_y = 0.82 + 0.19 * 1.15 = 1.04$ GPa at such pressure, which is contradictory.

## α-Zr

**Standard Rietveld refinement**

| $a/a_0$ | $c/c_0$ | $V/V_0$ |
|---|---|---|
| 0.996 | 0.995 | 0.988 |

**Pressure (GPa) based on hydrostatic generalized EoS**

| $p_a$ | $p_c$ | $p_v$ |
|---|---|---|
| 0.987 | 1.463 | 1.138 |

**In plane Rietveld refinement**

| $\bar{E}_{0,rr}$ | $\bar{E}_{0,\theta\theta}$ | $\bar{E}_{0,zz}$ | $\bar{E}_{0,v}$ |
|---|---|---|---|
| -0.0032 | -0.0033 | -0.0055 | -0.012 |

**Stresses (GPa) based on Hooke's law under pressure**

| $\sigma_{rr}$ | $\sigma_{\theta\theta}$ | $\sigma_{zz}$ | $p$ | $\sigma_{rr}-\sigma_{zz}$ | $\sigma_y(p)$ |
|---|---|---|---|---|---|
| -1.07 | -1.07 | -1.33 | 1.15 | 0.26 | 1.04 |

**CEA**

| $\bar{E}_{0,rr}$ | $\bar{E}_{0,\theta\theta}$ | $\bar{E}_{0,zz}$ | $\bar{E}_{0,v}$ |
|---|---|---|---|
| -0.00325 | -0.00325 | -0.01647 | -0.023 |

| $\sigma_{rr}$ | $\sigma_{\theta\theta}$ | $\sigma_{zz}$ | $p$ | $\sigma_{rr}-\sigma_{zz}$ | $\sigma_y(p)$ |
|---|---|---|---|---|---|
| -1.96 | -1.96 | -3.24 | 2.39 | 1.28 | 1.28 |

Figure S6: **Comparison of the stresses and elastic strains from the standard Rietveld refinement and CEA for $\alpha$-Zr.**

3. If alternatively, we use the same $\bar{E}_{0,rr} = \bar{E}_{0,\theta\theta} = -0.00325$ and determine $\bar{E}_{0,zz} = -0.01647$ from the developed CEA approach, we obtain $\sigma_{rr} = \sigma_{\theta\theta} = -1.96$ GPa, $\sigma_{zz} = -3.24$ GPa, and $p = 2.39$ GPa. The yield condition $|\sigma_{zz} - \sigma_{rr}| = \sigma_y = 0.82 + 0.19 * 2.38 = 1.28$ GPa is met, i.e., everything is consistent. The main reason for the difference of 1.24 GPa between $p$=1.15 GPa in the approach #2 and $p$=2.39 GPa, in which $\bar{E}_{0,zz}$ is determined with CEA to satisfy the yield condition, is that for axial XRD $\bar{E}_{0,zz}$ does not contribute to the XRD patterns, which is neglected in the traditional Rietveld refinement but is taken into account in the developed CEA method. Another conclusion is that standard Rietveld refinement for axial diffraction underestimates volumetric strain, -0.012 instead of -0.023.

Similar results for $\omega$-Zr are presented in Fig. S7. While difference in pressure between standard Rietveld refinement and that with CEA is significant (2.2 GPa), the pressure obtained with approach 2 is the same as with the CEA. However, stress components and especially their difference $|\sigma_{rr} - \sigma_{zz}|$ change significantly, by 1.9 GPa.

## 4 FEM simulations

### 4.1 Complete system of equations for FEM simulations [16]

Box 1 summarizes all equations derived in [16] in the form used in our simulations. Vectors and tensors are denoted in boldface type, e.g., $\boldsymbol{A} = A_{ij}\boldsymbol{e}_i\boldsymbol{e}_j$, where $A_{ij}$ are components in the Cartesian system with unit basis vectors $\boldsymbol{e}_i$ and summation over the repeated indices is assumed. Expressions $\boldsymbol{e}_i\boldsymbol{e}_j$ and $\boldsymbol{e}_i\boldsymbol{e}_k\boldsymbol{e}_t\boldsymbol{e}_d$ designate the direct or dyadic product of vectors, which represent second- and fourth-rank tensors, respectively. Let $\boldsymbol{A} \cdot \boldsymbol{B} = A_{ik}B_{kj}\boldsymbol{e}_i\boldsymbol{e}_j$ and $\boldsymbol{A} : \boldsymbol{B} = tr(\boldsymbol{A} \cdot \boldsymbol{B}) = A_{ij}B_{ji}$ be the contraction (or scalar product) of tensors over one

## $\boldsymbol{\omega}$ -Zr

**Standard Rietveld refinement**

| $a/a_0$ | $c/c_0$ | $V/V_0$ |
|---|---|---|
| 0.958 | 0.961 | 0.882 |

**Pressure (GPa) based on hydrostatic generalized EoS**

| $p_a$ | $p_c$ | $p_V$ |
|---|---|---|
| 15.2 | 15.5 | 15.3 |

**In plane Rietveld refinement**

| $E_{0,rr}$ | $E_{0,\theta\theta}$ | $E_{0,zz}$ | $E_{0,v}$ |
|---|---|---|---|
| -0.0364 | -0.0377 | -0.0439 | -0.118 |

**Stresses (GPa) based on Hooke's law under pressure**

| $\sigma_{rr}$ | $\sigma_{\theta\theta}$ | $\sigma_{zz}$ | $p$ | $\sigma_{rr}-\sigma_{zz}$ | $\sigma_y(p)$ |
|---|---|---|---|---|---|
| -17.1 | -17.1 | -18.3 | 17.5 | 1.2 | 3.1 |

**CEA**

| $E_{0,rr}$ | $E_{0,\theta\theta}$ | $E_{0,zz}$ | $E_{0,v}$ | $\sigma_{rr}$ | $\sigma_{\theta\theta}$ | $\sigma_{zz}$ | $p$ | $\sigma_{rr}-\sigma_{zz}$ | $\sigma_y(p)$ |
|---|---|---|---|---|---|---|---|---|---|
| -0.03705 | -0.03705 | -0.06102 | -0.135 | -16.5 | -16.5 | -19.6 | 17.5 | 3.1 | 3.1 |

Figure S7: **Comparison of the stresses and elastic strains from the standard Rietveld refinement and CEA for $\omega$ -Zr.**

and two nearest indices, where $tr$ is the trace operation (sum of the diagonal components), and $A_{ik}B_{kj}$ is the matrix product. In the equations, first $\cdot$ is performed, and then $:$, e. g., $\boldsymbol{A}:\boldsymbol{B}\cdot\boldsymbol{K}=\boldsymbol{A}:(\boldsymbol{B}\cdot\boldsymbol{K})$. The direct (or dyadic) product of two tensors $\boldsymbol{K}$ and $\boldsymbol{M}$ is the tensor $\boldsymbol{KM}$ of rank equal to the sum of the two initial ranks. In particular, for the second-rank tensors $\boldsymbol{K}=K_{ij}\boldsymbol{e}_i\boldsymbol{e}_j$ and $\boldsymbol{M}=M_{kl}\boldsymbol{e}_k\boldsymbol{e}_l$, one has $\boldsymbol{KM}=K_{ij}M_{kl}\boldsymbol{e}_i\boldsymbol{e}_j\boldsymbol{e}_k\boldsymbol{e}_l$. Also, $\boldsymbol{A}_s=:\frac{\boldsymbol{A}+\boldsymbol{A}^t}{2}$ and $\boldsymbol{A}_a=:\frac{\boldsymbol{A}-\boldsymbol{A}^t}{2}$ are respectively the symmetric and anti-symmetric components of $\boldsymbol{A}$, where $'t'$ in the superscript designates the transpose operation defined as $\boldsymbol{A}^t=A_{ji}\boldsymbol{e}_i\boldsymbol{e}_j$, when $\boldsymbol{A}=A_{ij}\boldsymbol{e}_i\boldsymbol{e}_j$.

**Box 1. Complete system of equations**

*Decomposition of the total deformation gradient $\boldsymbol{F}$ into elastic $\boldsymbol{F}_e$ and inelastic $\boldsymbol{F}_i$ contributions:*

$$\boldsymbol{F} = \frac{\partial \boldsymbol{r}}{\partial \boldsymbol{r}_0} = \boldsymbol{V}_e \cdot \boldsymbol{R}_e \cdot \boldsymbol{U}_i = \boldsymbol{V}_e \cdot \bar{\boldsymbol{F}}_i; \bar{\boldsymbol{F}}_i = \boldsymbol{R}_e \cdot \boldsymbol{U}_i; \boldsymbol{B}_e = 0.5(\boldsymbol{F}_e \cdot \boldsymbol{F}_e^t - \boldsymbol{I}) = 0.5(\boldsymbol{V}_e^2 - \boldsymbol{I}),$$
(S.90)

where $\boldsymbol{r}$ and $\boldsymbol{r}_0$ are respectively the position vectors of the material point in current and reference configurations. $\boldsymbol{V}_e$ and $\boldsymbol{R}_e$ are from the polar decomposition of $\boldsymbol{F}_e$; $\boldsymbol{U}_i$ is the right stretch tensor from the polar decomposition of $\boldsymbol{F}_i$ and $\boldsymbol{B}_e$ is the elastic Eulerian strain tensor. Decomposition of the deformation rate $\boldsymbol{d}$ into elastic, plastic, and transformation

parts:

$$\boldsymbol{d} = \overset{\nabla}{\boldsymbol{B}}_e \cdot \boldsymbol{V}_e^{-2} + 2(\boldsymbol{d} \cdot \boldsymbol{B}_e)_a \cdot \boldsymbol{V}_e^{-2} + \boldsymbol{\gamma} + \bar{\varepsilon}_{t0}\dot{c}\boldsymbol{I} \tag{S.91}$$

where $\overset{\nabla}{\boldsymbol{B}}_e = \dot{\boldsymbol{B}}_e - 2(\boldsymbol{w} \cdot \boldsymbol{B}_e)_s$ is the Jaumann time derivative of $\boldsymbol{B}_e$, $\boldsymbol{w}$ is the anti-symmetric part of the velocity gradient in the current configuration $\boldsymbol{l}$; $\boldsymbol{\gamma}$ is the plastic part of the deformation rate, $\bar{\varepsilon}_{t0}$ is the volumetric transformation strain, and $\boldsymbol{I}$ is the unit tensor.

*The third -order Murnaghan potential:*

$$\psi(\boldsymbol{B}_e) = \frac{\lambda + 2G}{2}I_1^2 - 2GI_2 + \left(\frac{l+2m}{3}I_1^3 - 2mI_1I_2 + nI_3\right), \tag{S.92}$$

where $\lambda$, $G$, $l$, $m$, and $n$ are the Murnaghan material constants of the mixture and $I_1$, $I_2$, and $I_3$ are the invariants of $\boldsymbol{B}_e$ defined as

$$\begin{aligned}I_1 &= B_{e11} + B_{e22} + B_{e33}; \\ I_2 &= B_{e22}B_{e33} - B_{e23}^2 + B_{e11}B_{e33} - B_{e13}^2 + B_{e22}B_{e11} - B_{e12}^2; \quad I_3 = det\boldsymbol{B}_e.\end{aligned} \tag{S.93}$$

Simple mixture rule is used to obtain the Murnaghan constants of the mixture

$$\begin{aligned}\lambda &= (1-c)\lambda_1 + c\lambda_2; \quad G = (1-c)G_1 + cG_2; \quad m = (1-c)m_1 + cm_2; \\ l &= (1-c)l_1 + cl_2; \quad n = (1-c)n_1 + cn_2.\end{aligned} \tag{S.94}$$

Here, the subscripts 1 and 2 designate $\alpha$- and $\omega$-Zr, c is the concentration of the $\omega$-Zr.

*Elasticity rule for the Cauchy (true) stress:*

$$\begin{aligned}\boldsymbol{\sigma} &= J_e^{-1}(2\boldsymbol{B}_e + \boldsymbol{I}) \cdot \frac{\partial \psi}{\partial \boldsymbol{B}_e} \\ &= J_e^{-1}(2\boldsymbol{B}_e + \boldsymbol{I}) \cdot \left(\lambda I_1 \boldsymbol{I} + 2G\boldsymbol{B}_e + (lI_1^2 - 2mI_2)\boldsymbol{I} + n\frac{\partial I_3}{\partial \boldsymbol{B}_e} + 2mI_1\boldsymbol{B}_e\right),\end{aligned} \tag{S.95}$$

where $J_e = det\boldsymbol{F}_e$ is the Jacobian determinant of $\boldsymbol{F}_e$. Compact expressions of $\frac{\partial I_1}{\partial \boldsymbol{B}_e}$, $\frac{\partial I_2}{\partial \boldsymbol{B}_e}$, and $\frac{\partial I_3}{\partial \boldsymbol{B}_e}$ are

$$\frac{\partial I_1}{\partial \boldsymbol{B}_e} = \boldsymbol{I}; \quad \frac{\partial I_2}{\partial \boldsymbol{B}_e} = -\boldsymbol{B}_e + I_1\boldsymbol{I}; \quad \frac{\partial I_3}{\partial \boldsymbol{B}_e} = \boldsymbol{B}_e \cdot \boldsymbol{B}_e - I_1\boldsymbol{B}_e + I_2\boldsymbol{I}. \tag{S.96}$$

*Yield surface:*

$$\phi = \sqrt{3/2\boldsymbol{s}:\boldsymbol{s}} - (\sigma_{y0} + bp) = 0; \sigma_{y0} = (1-c)\sigma_{y01} + c\sigma_{y02}; b = (1-c)b_1 + cb_2. \tag{S.97}$$

Here, $\boldsymbol{s}$ is the deviatoric part of the Cauchy stress $\boldsymbol{\sigma}$, $p$ is the pressure, $\sigma_{y01}$ and $\sigma_{y02}$ are the yield strengths in compression of the $\alpha$- and $\omega$-Zr @$p = 0$, respectively, $b_1$ and $b_2$ are their linear pressure hardening coefficients.

*Plastic flow rule:*

$$\boldsymbol{\gamma} = |\boldsymbol{\gamma}|\frac{\boldsymbol{s}}{\sqrt{\boldsymbol{s}:\boldsymbol{s}}} = |\boldsymbol{\gamma}|\boldsymbol{n}; |\boldsymbol{\gamma}| = (\boldsymbol{\gamma}:\boldsymbol{\gamma})^{0.5} \text{ when } \phi(\boldsymbol{s},p,c) = 0 \text{ and } \dot{\phi}(\boldsymbol{s},p,c) = 0, \qquad \text{(S.98)}$$

i.e., in the elastoplastic region, and $|\boldsymbol{\gamma}|$ is determined from the consistency condition $\dot{\phi}(\boldsymbol{s},p,c) = 0$; where $|\boldsymbol{\gamma}| = 0$ in the elastic region when $\phi(\boldsymbol{s},p,c) < 0$ or $\phi(\boldsymbol{s},p,c) = 0$ and $\dot{\phi}(\boldsymbol{s},p,c) = 0$.

$$|\boldsymbol{\gamma}| = \frac{-\left(\frac{\sqrt{1.5}\boldsymbol{s}}{\sqrt{\boldsymbol{s}:\boldsymbol{s}}} + \frac{b}{3}\boldsymbol{I}\right):\boldsymbol{Y}}{\left(\frac{\sqrt{1.5}\boldsymbol{s}}{\sqrt{\boldsymbol{s}:\boldsymbol{s}}} + \frac{b}{3}\boldsymbol{I}\right):\boldsymbol{Z} + \frac{\partial \phi}{\partial c}A\sqrt{\frac{2}{3}}} : \boldsymbol{d} \qquad \text{(S.99)}$$

where $\boldsymbol{Y} = \frac{\partial \boldsymbol{\sigma}}{\partial \boldsymbol{B}_e} \cdot \boldsymbol{V}_e^2$, $\boldsymbol{Z} = -\boldsymbol{Y}:\left[\boldsymbol{n} + (\bar{\varepsilon}_{t0}\boldsymbol{I} + \boldsymbol{\gamma}_t)A\sqrt{\frac{2}{3}} + \frac{\partial f}{\partial c}A\sqrt{\frac{2}{3}}\right]$, $\boldsymbol{\sigma} = \boldsymbol{f}(\boldsymbol{B}_e, c)$, and $\frac{dc}{dq} = A(p,q,c)$.

*Accumulated plastic strain:*

$$\dot{q} = \sqrt{2/3}|\boldsymbol{\gamma}| \qquad \text{(S.100)}$$

*Stress rate – deformation rate relationship:*

$$\overset{\triangledown}{\boldsymbol{\sigma}} = \left(\boldsymbol{Y} + \boldsymbol{Z}\frac{-\left(\frac{\sqrt{1.5}\boldsymbol{s}}{\sqrt{\boldsymbol{s}:\boldsymbol{s}}} + \frac{b}{3}\boldsymbol{I}\right):\boldsymbol{Y}}{\frac{\partial \phi}{\partial \boldsymbol{\sigma}}:\boldsymbol{Z} + \left(\frac{\partial \phi}{\partial c}A + \frac{\partial \phi}{\partial q}\right)\sqrt{\frac{2}{3}}}\right):\boldsymbol{d} \qquad \text{(S.101)}$$

*Equilibrium equation:*

$$\boldsymbol{\nabla} \cdot \boldsymbol{\sigma} = \boldsymbol{0} \qquad \text{(S.102)}$$

### 4.2 Geometry and boundary conditions

Geometry of DAC is shown in see Fig. 1a in the main text. Axisymmetric problem formulation is considered. Geometry of the sample and the anvil, as well as the boundary conditions, are shown in see Fig. S8. They are:

(1) A uniform vertical displacement is applied at the boundary between the top inclined surface of the anvil and Boehler-type seat (line CD). Distributions of stresses or displacements along this surface do not affect fields in the sample and the diamond close to the diamond culet.

(2) At the symmetry axis $r = 0$ (line AB), $\tau_{rz}$ and horizontal displacement are zero. At the symmetry plane $z = 0$, shear stress $\tau_{rz}$ and vertical displacement are zero.

(3) At the contact surface between the sample and the anvil, an isotropic friction model, described below, is utilized.

(4) Other surfaces not mentioned above are stress-free. Quadrilateral 4-node bilinear axisymmetric finite elements CGAX4R are used in simulations, which are commonly used for large-deformation axisymmetric problems [17]. Our simulations utilize a mesh with 4271 elements.

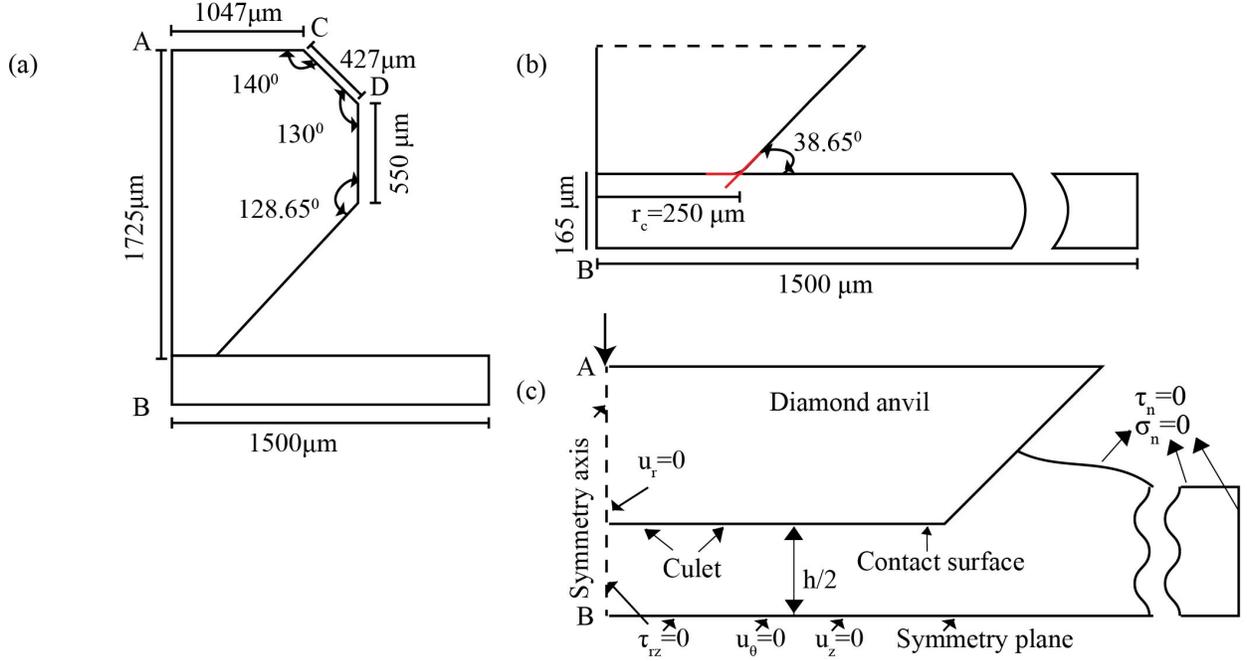

Figure S8: **Details of the geometrical model of the DAC experiment used in FEM:** (a) Description of the one half of diamond dimensions in the natural configuration. (b) Geometric parameters of one half of sample and the angle between culet and non-culet diamond surfaces. (c) Boundary conditions and the schematic of sample in the deformed configuration.

Evolution of concentration of $\omega$ phase $c(r)$ and corresponding volumetric transformation strain $\bar{\varepsilon}_{t0} \dot{c} \boldsymbol{I}$ ($\bar{\varepsilon}_{t0} = -0.0158$) are introduced homogeneously along the z-coordinate in our FEM problem formulation for each loading step.

**Friction model:**

At the culet portion of the diamond $r \leq r_c$, the contact shear stress is given by

$$\tau_c = m' \tau_y(p) \qquad \text{for } r \leq r_c \tag{S.103}$$

The distribution of $m'(r)$ is obtained from the analytical solution (Fig. S9).

At the inclined portion of the sample-diamond contact surface, the critical shear is governed by the combined Coulomb friction $\tau_{cr} = \mu(\sigma_c)\sigma_c$ and Eq. (S.103), where $\sigma_c$ is the contact normal stress. There is complete cohesion between the sample and the anvil unless shear stress $\tau_c$ reaches the critical value:

$$\tau_c < \tau_{cr} = min\left[\mu(\sigma_c)\sigma_c, m'(r_c)\tau_y(p)\right] \rightarrow cohesion, \tag{S.104}$$

where $m' = m'(r_c)$ is determined at $r = r_c$. When friction stress reaches $\tau_{cr}$,

$$\tau_c = \tau_{cr} = min[\mu(\sigma_c)\sigma_c, m'(r_c)\tau_y(p)] \rightarrow sliding, \tag{S.105}$$

contact sliding occurs.

### 4.3 Nonlinear elastic equations and material properties for single-crystal diamond anvil

The constitutive response of diamond is modeled using fourth-order nonlinear anisotropic elastic potential energy given by Vekilov et al. [18]:

$$\begin{aligned}\psi =& \frac{1}{2}c_{11}(\eta_1^2+\eta_2^2+\eta_3^2)+c_{12}(\eta_1\eta_2+\eta_2\eta_3+\eta_1\eta_3)+\frac{1}{2}c_{44}(\eta_4^2+\eta_5^2+\eta_6^2)+\frac{1}{6}c_{111}(\eta_1^3+\eta_2^3+\\ & \eta_3^3)+\frac{1}{2}c_{112}[\eta_1^2(\eta_2+\eta_3)+\eta_2^2(\eta_1+\eta_3)+\eta_3^2(\eta_1+\eta_2)]+c_{123}\eta_1\eta_2\eta_3+c_{456}\eta_4\eta_5\eta_6+\\ & \frac{1}{2}c_{144}(\eta_1\eta_4^2+\eta_2\eta_5^2+\eta_3\eta_6^2)+\frac{1}{2}c_{155}[\eta_4^2(\eta_2+\eta_3)+\eta_5^2(\eta_1+\eta_3)+\eta_6^2(\eta_1+\eta_2)]+\\ & \frac{1}{24}c_{1111}(\eta_1^4+\eta_2^4+\eta_3^4)+\frac{1}{6}c_{1112}[\eta_1^3(\eta_2+\eta_3)+\eta_2^3(\eta_1+\eta_3)+\eta_3^3(\eta_1+\eta_2)]+\frac{1}{4}c_{1144}(\eta_1^2\eta_4^2+\\ & \eta_2^2\eta_5^2+\eta_3^2\eta_6^2)+\frac{1}{4}c_{1122}(\eta_1^2\eta_2^2+\eta_2^2\eta_3^2+\eta_3^2\eta_1^2)+\frac{1}{2}c_{1123}\eta_1\eta_2\eta_3(\eta_1+\eta_2+\eta_3)+\\ & \frac{1}{4}c_{1155}[\eta_1^2(\eta_6^2+\eta_5^2)+\eta_2^2(\eta_6^2+\eta_4^2)+\eta_3^2(\eta_5^2+\eta_4^2)]+\frac{1}{2}c_{1255}[\eta_1\eta_2(\eta_4^2+\eta_5^2)+\eta_3\eta_2(\eta_6^2+\\ & \eta_5^2)+\eta_1\eta_3(\eta_6^2+\eta_4^2)]+\frac{1}{2}c_{1266}(\eta_1\eta_2\eta_6^2+\eta_2\eta_3\eta_4^2+\eta_1\eta_3\eta_5^2)+c_{1456}\eta_4\eta_5\eta_6(\eta_1+\eta_2+\eta_3)+\\ & \frac{1}{24}c_{4444}(\eta_4^4+\eta_5^4+\eta_6^4)+\frac{1}{4}c_{4455}(\eta_4^2\eta_5^2+\eta_6^2\eta_5^2+\eta_4^2\eta_6^2),\end{aligned}$$
(S.106)

where $\eta_1 = E_{e11}$, $\eta_2 = E_{e22}$, $\eta_3 = E_{e33}$, $\eta_4 = 2E_{e23}$, $\eta_5 = 2E_{e31}$, and $\eta_6 = 2E_{e12}$ are the Lagrangian strains. Based on the elasticity law, the Cauchy stress in the diamond can be obtained using:

$$\boldsymbol{\sigma} = \frac{1}{J}\boldsymbol{F}_e \cdot \frac{\partial\psi}{\partial \boldsymbol{E}_e} \cdot \boldsymbol{F}_e^t.$$
(S.107)

Here $J$ is the Jacobian determinant of $\boldsymbol{F}_e$. All the elastic constants of diamond are taken from Telichko et al. [19] and they are as follows (all in GPa):

$c_{11} = 1081.9, c_{12} = 125.2, c_{44} = 578.6;$
$c_{111} = -7611, c_{112} = -1637, c_{123} = 604, c_{144} = -199, c_{166} = -2799, c_{155} = -2799,$
$c_{456} = -1148, c_{1111} = 26687, c_{1112} = 9459, c_{1122} = 6074, c_{1123} = -425, c_{1144} = -1385,$
$c_{1155} = 10741, c_{1255} = -264, c_{1266} = 8192, c_{1456} = 487, c_{4444} = 11328, c_{4455} = 528.$
(S.108)

### 4.4 Elastic properties of polycrystalline $\alpha$- and $\omega$-Zr:

The elastic constitutive response of polycrystalline Zr is modeled using the third-order nonlinear Murnaghan potential Eq. (S.92). Out of 5 elastic constants in the Murnaghan potential, 2 elastic constants, Lame constant $\lambda$ and shear modulus $G$, are related to the quadratic in $\boldsymbol{B}_e$ terms, and the rest, $l$, $m$ and $n$, are related to the cubic in $\boldsymbol{B}_e$ terms. These constants are calibrated using the bulk modulus $K$ and its pressure derivative $\frac{dK}{dp}@p=0$ obtained from the pressure-volume relationships in hydrostatic DAC experiments, and the shear modulus $G$ and its pressure-derivative $\frac{dG}{dp}@p=0$ are taken from the experimental results [20, 21]. The expressions relating the Murnaghan constants $\lambda$, $G$, $l$, $m$, $n$ and $K$, $G$, $\frac{dK}{dp}$ and $\frac{dG}{dp}$ @ $p=0$ are:

$$K = \frac{3\lambda + 2G}{3}; \quad \frac{dK}{dp} = K' = -\frac{2(9l+n)}{9K}; \quad \frac{dG}{dp} = G' = \frac{-2G - 6K - 6m + n}{6K}. \quad \text{(S.109)}$$

It can be seen there are only 2 equations to solve for the 3 third-order constants. Therefore, there is an indeterminacy of degree 1. However, it can be easily shown for any pressure, that when the deviatoric part of the superposed deformation is small, the stresses and energy can be written in terms of just $K$, $G$, $K'$, $G'$. Therefore, one of the constants, $l$, $m$, or $n$, can be chosen arbitrary, and two other are determined from Eq. (S.109). The constants that are used are (all in GPa):

$$\lambda = 68.11, \quad G = 36.13, \quad l = -147.01, m = -122.75, \quad n = -100 \quad \text{for } \alpha - Zr;$$
$$\lambda = 72.33, \quad G = 45.1, \quad l = -149.56, m = -179.53, \quad n = -4 \quad \text{for } \omega - Zr.$$
$$\text{(S.110)}$$

**Friction stress and rules for $\alpha$- and $\omega$-Zr and their mixture:**

Friction along the sample-diamond contact surface in DAC experiment plays the central role in generating high pressure in the sample. However, the magnitude of the friction force along the contact surface cannot be measured. Traditionally [13, 22, 23], the yield strength in shear was assumed to be equal to the sliding friction force. But it was reported recently in [8, 24] that the sliding friction force is significantly lower than the yield strength in shear. This important deduction, coupled with the elusive nature of the mechanism of the sliding friction at high pressure, posed an important problem to find boundary conditions for FEM solution and increased the degree of indeterminacy of all fields in a sample. The problem is resolved by using the results of analytical model that gives the distributions of contact shear stress (Fig. S9). For the mixture of $\alpha+\omega$ Zr, there is a jump in yield strength in shear of mixture, which is due to the jump in the concentration of the $\omega$-Zr phase (See Fig. 2 in main text).

To find friction sliding laws, we must exclude regions near the symmetry axis. At the symmetry axis, friction stress is zero due to symmetry, which leads to zero or very small sliding near the symmetry. We also exclude region near the culet edge, where character of the plastic flow significantly changes.

The best linear fits for the contact friction stress $\tau_c$ normalized by the yield strength in shear $\tau_y^c$ at the same point for pure-$\omega$ Zr in the region $r = 60$ $\mu m$ to 140 $\mu m$, for the mixture of $\alpha$ and $\omega$ Zr in the region r=130 $\mu m$ to 200 $\mu m$, and for the pure-$\alpha$ Zr in the region r=130

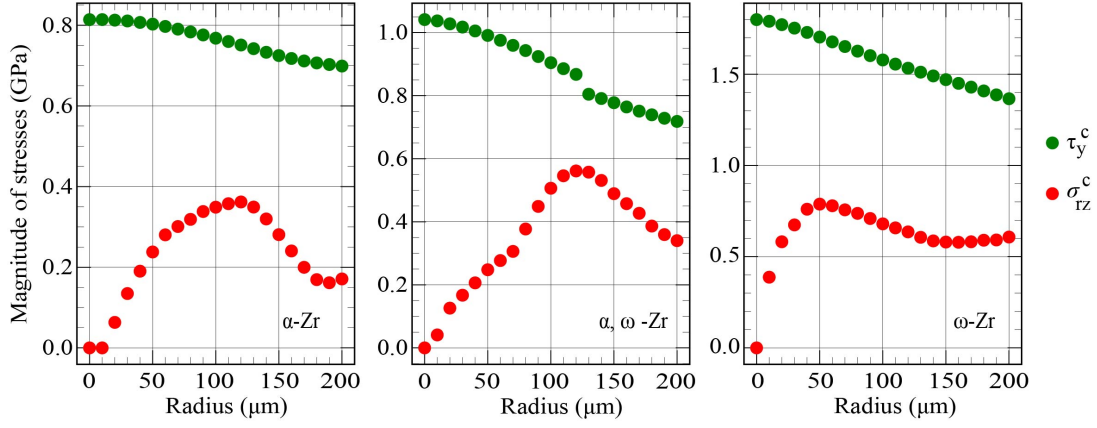

Figure S9: **Radial distributions of shear stress and yield strength in shear for** pure -Zr (a), mixture of $\alpha$- and $\omega$-Zr (b), and pure $\omega$-Zr (c) obtained using analytical model and experimentally measured radial $\bar{E}_{0,rr}$ and azimuthal $\bar{E}_{0,\theta\theta}$ strain distributions.

$\mu m$ to 190 $\mu m$ are (Fig. S10):

$$\left(\frac{\tau_c}{\tau_y^c}\right)_\omega = 0.186 + 0.018 p^c \quad \text{for } 11.1 \leq p^c(GPa) \leq 15.0; \quad 60 \leq r(\mu m) \leq 140;$$

$$\left(\frac{\tau_c}{\tau_y^c}\right)_{\alpha+\omega} = 0.179 + 0.241 p^c \quad \text{for } 2.7 \leq p^c(GPa) \leq 3.7; \quad 130 \leq r(\mu m) \leq 200; \quad \text{(S.111)}$$

$$\left(\frac{\tau_c}{\tau_y^c}\right)_\alpha = -1.282 + 0.722 p^c \quad \text{for } 2.0 \leq p^c(GPa) \leq 2.45; \quad 130 \leq r(\mu m) \leq 190;$$

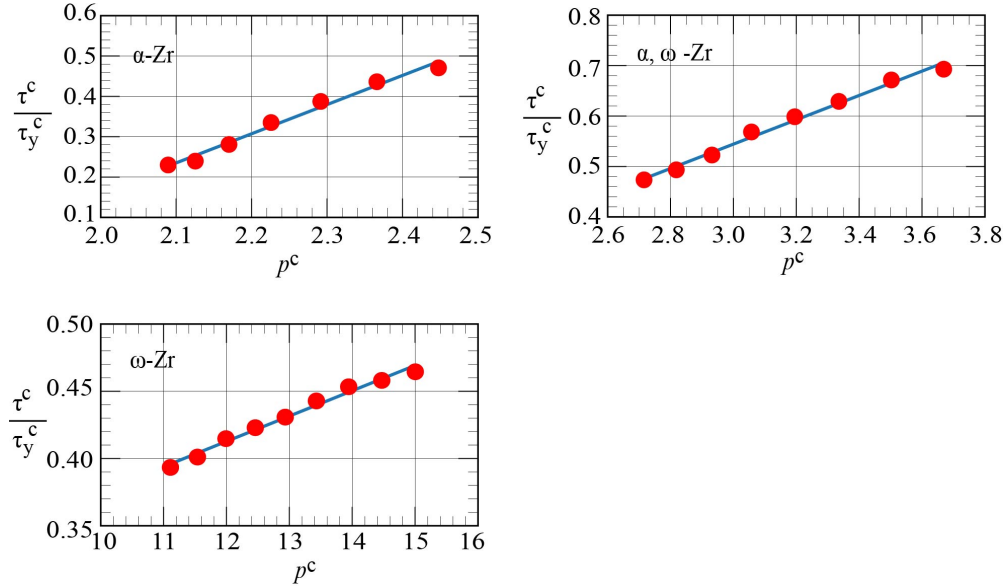

Figure S10: **Pressure dependence of the contact friction stress $\tau_c$ normalized by the yield strength in shear $\tau_y^c$ at the same point.** Results are presented for pure $\alpha$-Zr at $p_{max} = 3.09$ GPa (a), mixture of $\alpha$- and $\omega$-Zr for $p_{max} = 4.77$ GPa (b), and pure $\omega$-Zr at $p_{max} = 17.55$ GPa (c).

Supplementary figures

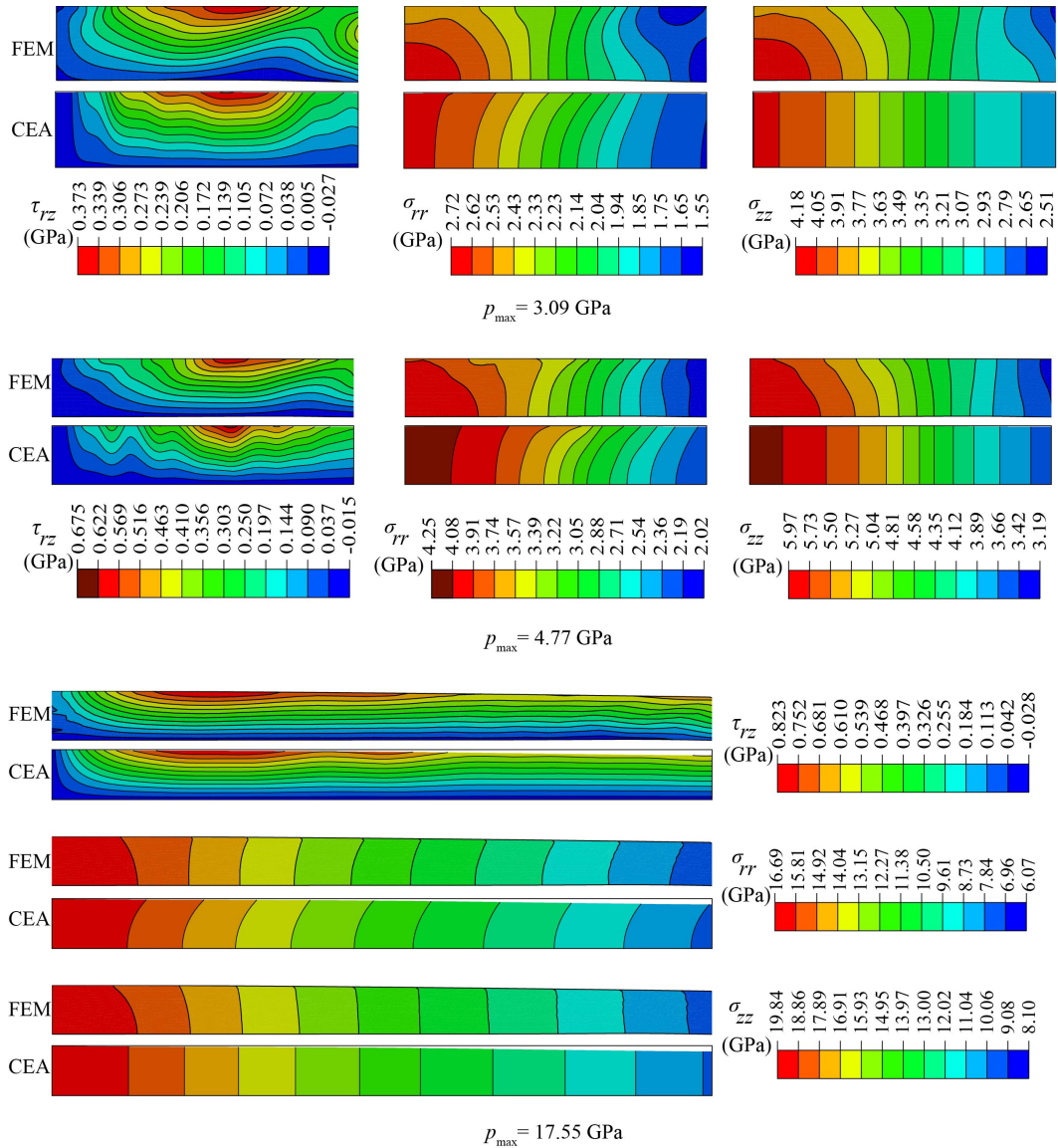

Figure S11: **2D stress contours for three loading cases from FEM (top) and analytical (bottom) solutions.**

2D stress contours for 3 loading cases Fig. S11.

**References**

[1] A.P. Zhilyaev, F. Gálvez, A. Sharafutdinov, M.T. Pérez-Prado, Influence of the high-pressure torsion die geometry on the allotropic phase transformations in pure Zr. Mat. Sci. Eng. A 527, 3918-3928 (2010).

[2] A. P. Hammersley, Fit2d: An introduction and overview, in ESRF Internal Report, ESRF97HA02T (Institute of Physics, Bristol, 1997).